\documentclass[longbibliography,singlecolumn,showpacs,nofootinbib,superscriptaddress,notitlepage]{revtex4-1}
\pdfoutput=1
\usepackage{amsmath}
\usepackage{amssymb,bm}
\usepackage{amsthm}
\usepackage[utf8]{inputenc}
\usepackage[english]{babel}
\usepackage[T1]{fontenc}
\usepackage{tikz}

\newtheorem{innercustomgeneric}{\customgenericname}
\providecommand{\customgenericname}{}
\newcommand{\newcustomtheorem}[2]{%
  \newenvironment{#1}[1]
  {%
   \renewcommand\customgenericname{#2}%
   \renewcommand\theinnercustomgeneric{##1}%
   \innercustomgeneric
  }
  {\endinnercustomgeneric}
}

\newcustomtheorem{customthm}{Theorem}
\newcustomtheorem{definitionBob}{Definition}

\usepackage{color,dsfont} 
\usepackage{graphicx}
\usepackage{ragged2e}
\usepackage[colorlinks=true, hyperindex, breaklinks, linkcolor=blue, urlcolor=blue, citecolor=blue]{hyperref} 
\usepackage[normalem]{ulem}
\usepackage[capitalise]{cleveref}
\usepackage{mathrsfs}
\usepackage[caption=false]{subfig}
\usepackage{mathtools}
\usepackage{float}
\usepackage{verbatim}
\usepackage{latexsym}
\usepackage{amsmath}
\usepackage{amssymb}
\usepackage{setspace}
\usepackage{amsfonts}
\usepackage{stmaryrd}
\usepackage{xcolor}
\usepackage{enumitem}
\usepackage{lipsum}

\DeclareMathOperator*{\argmax}{arg\,max}

 \newtheorem{definition}{Definition}

\newcommand{\bra}[1]{\langle#1|}  
\newcommand{\ket}[1]{|#1\rangle}  
 
\newcommand{\License}[1]{\begingroup
  \renewcommand\thefootnote{}
  \footnote{#1}%
  \addtocounter{footnote}{-1}%
  \endgroup
}

\begin{document}

\title{Fault-tolerant preparation of approximate GKP states}

\author{Yunong Shi}
\email{yunong@uchicago.edu}
\affiliation{
   Physics department, University of Chicago, Chicago, IL 60637, United States
    }

\author{Christopher Chamberland}
\email{christopher.chamberland@ibm.com}
\affiliation{
   IBM T. J. Watson Research Center,
    Yorktown Heights, NY, 10598, United States
    }
    
    \author{Andrew Cross}
\email{awcross@us.ibm.com}
\affiliation{
   IBM T. J. Watson Research Center,
    Yorktown Heights, NY, 10598, United States
    }
\License{Y.S. and C.C. contributed equally to the work.}

\begin{abstract}

Gottesman-Kitaev-Preskill (GKP) states appear to be amongst the leading candidates for correcting errors when encoding qubits into oscillators. However the preparation of GKP states remains a significant theoretical and experimental challenge. Until now, no clear definitions for fault-tolerantly preparing GKP states have been provided. Without careful consideration, a small number of faults can lead to large uncorrectable shift errors. After proposing a metric to compare approximate GKP states, we provide rigorous definitions of fault-tolerance and introduce a fault-tolerant phase estimation protocol for preparing such states. The fault-tolerant protocol uses one flag qubit and accepts only a subset of states in order to prevent measurement readout errors from causing large shift errors. We then show how the protocol can be implemented using circuit QED. In doing so, we derive analytic expressions which describe the leading order effects of the non-linear dispersive shift and Kerr non-linearity. Using these expressions, we show that to mitigate the non-linear dispersive shift and Kerr terms would require the protocol to be implemented on time scales four orders of magnitude longer than the time scales relevant to the protocol for physically motivated parameters. Despite these restrictions, we numerically show that a subset of the accepted states of the fault-tolerant phase estimation protocol maintain good error correcting capabilities even in the presence of noise. 

\end{abstract}


\maketitle

\section{Introduction}
\label{sec:Intro}

Fault-tolerant quantum computing will be essential for implementing large scale quantum algorithms that offer provable speed-ups over the best known classical algorithms. Currently there are many proposals for encoding qubits into error correcting codes in order to perform universal fault-tolerant quantum computation. Depending on the underlying physical architecture, some encoding schemes are more suitable than others. 

One method proposed by Gottesman, Kitaev and Preskill is to encode a qubit into an oscillator such that small shift errors in both position and momentum can be corrected. Although some bosonic codes have been designed to correct realistic errors arising from noise models encountered in the experiment (e.g. photon loss), recently it has been shown that GKP codes have better error correction capabilities than such codes under the assumption of perfect encoding and decoding \cite{AlbertBosonicCodePerformance18,GaussianChannel18,BarbaraDanielGKP}. In addition, it has been shown how GKP codes can be concatenated with the toric code in order to achieve larger threshold values compared to toric codes with bare physical qubits  \cite{ToricGKPBarbara,GKPFaultToleranceJapan,NohChambsGKPSurface}. Lastly, given a supply of GKP-encoded Pauli eigenstates, universal fault-tolerant quantum computation can be achieved using only Gaussian operations \cite{MagicStateGKP19}.

Given the above, it is clear that the fault-tolerant preparation of encoded GKP states is an important problem that needs to be addressed. Various proposals for preparing GKP states have been outlined \cite{FirstGKPPrep02,GlancyKnillCondition,BarbaraDanielGKP,GKPStatePrepHome16,HomeNatureGKP,TaiwanGKPPrep18,GKPStatePrepHomePhotonCatalysis,ProbFaultGKP19,ShrutiStabilizedCat19}. However to our knowledge, no clear definitions for fault-tolerantly preparing GKP states using qubit-cavity couplings have been proposed. As such, without careful consideration, it is possible that a small number of faults lead to large uncorrectable shift errors. Inspired by \cite{AGP06}, in this work we propose new fault-tolerant definitions for preparing GKP states which tolerate small shift errors and a small number of faults occurring on ancilla qubits during the protocol. We then show how the phase estimation protocol proposed in this work satisfies our fault-tolerance criteria. In particular, the protocol is robust to a single fault occurring on the ancilla qubits in addition to shift errors of magnitude at most $0.06$. In order to be fault-tolerant, the protocol uses one flag qubit (and thus requires a total of two ancilla qubits) to prevent damping errors and imperfect implementations of the required gates from causing large shift errors. In addition, we provide an algorithm which only accepts a subset of all output states of phase estimation in order to prevent a single measurement readout error from causing large uncorrectable shift errors. We then proceed to show how our protocol can be implemented using circuit QED. We first analytically derive expressions describing the effects of the non-linear dispersive shift and Kerr non-linearity on the evolution of the cavity. We then numerically show that certain states output from the phase estimation protocol are robust to noise processes found in current 2D and 3D cavities since these can still correct small shift errors with high probability.

Our paper is structured as follows. In \cref{sec:GoodnessGKP}, we provide new metrics which we use throughout the remainder of the paper to characterize the error correction capabilities of approximate GKP states. The fault-tolerance definitions used throughout this paper are given in \cref{sec:FaultToltDefs}. In \cref{subsec:FaulFreePhaseEstimation}, we briefly review the phase estimation protocol described in \cite{BarbaraDanielGKP}. In \cref{subsec:FTPhaseEstOneDelt}, we obtain a new phase estimation protocol and prove that it is fault-tolerant under our definitions. In \cref{sec:GoodnessNoiseFreeSection}, we compare the error correction capabilities of various states obtained from the phase estimation protocol in the noise free case. \cref{sec:CircuitQED} is devoted to the implementation and error analysis of our protocol in circuit QED. In \cref{subsec:KerrDeriv}, we provide analytic expressions for the time evolution of the qubit-cavity coupling when implementing a controlled-displacement gate. The expressions are derived in \cref{App:NoiseAnalysisSection}. In \cref{subsec:MasterEquationAnalysis}, we numerically solve a master equation which includes all considered noise processes, such qubit damping and dephasing, photon loss in addition to measurement, ancilla state-preparation and gate errors which arise from a depolarizing noise channel. In \cref{sec:conclusion} we summarize our results and discuss possible future directions.

\section{Goodness of approximate GKP states}
\label{sec:GoodnessGKP}

As was explained in \cite{GKP2001,BarbaraDanielGKP}, preparing perfect GKP states would require an infinite amount of squeezing. In a realistic setting, one can only prepare approximate GKP states with finite squeezing. Perfect GKP states, which are $+1$ eigenstates of the mutually commuting operators $S_{p} = e^{-2i \sqrt{\pi}p}$ and $S_{q} = e^{2i \sqrt{\pi} q}$, can correct shift errors of size at most $\frac{\sqrt{\pi}}{2}$.  Note that for the displacement operator $D(\alpha) = e^{\alpha a^{\dagger} - \alpha^{*} a}$, we can write $S_p = D(\sqrt{2 \pi})$ and $S_q =  D(i \sqrt{2 \pi})$. The goal of this paper will be to fault-tolerantly prepare approximate GKP states which can correct small shift errors correctable by perfect GKP states with high probability (in \cref{sec:FaultToltDefs} we will specify what we mean by correctable shift errors). Therefore, it is important to have a metric which allows us to compute the "goodness" of an approximate GKP state. 

Recall that for perfect GKP states, the logical $\ket{\overline{0}}$ and $\ket{\overline{1}}$ states are given by
\begin{align}
\ket{\overline{0}} = \sum_{k= - \infty}^{\infty} S_{p}^{k} \ket{q=0} \nonumber \\
\ket{\overline{1}} =\sum_{k= - \infty}^{\infty} S_{p}^{k} \ket{q=\sqrt{\pi}},
\label{eq:ExactGKPstates}
\end{align}
up to normalization. In practice, approximate GKP states analogous to those in \cref{eq:ExactGKPstates} can be prepared by first preparing finitely squeezed states in $q$ space and approximately projecting these states onto the $S_{p} = 1$ eigenspace. For instance, the $\ket{q = 0}$ state can be written as a squeezed vacuum state $\ket{\text{sq}}$ with squeezing parameter $\Delta$ which is given by
\begin{align}
\ket{\text{sq}} = \int \frac{dq}{(\pi \Delta^{2})^{1/4}} e^{-q^{2}/ (2 \Delta^2)} \ket{q}.
\label{eq:SqueezedVacuumState}
\end{align}
One can then apply a sum of displacements with a Gaussian filter to obtain
\begin{align}
\ket{\overline{0}}_{\text{approx}} = C \sum _{k = - \infty}^{\infty} e^{-2 \pi \tilde{\Delta}^{2} k^{2}}D(k \sqrt{2 \pi}) \ket{\text{sq}},
\end{align}
where $C$ is a normalization coefficient. As was shown in \cite{BarbaraDanielGKP}, it is natural to have $\tilde{\Delta} = \Delta$. In \cref{Sec:FaultTolerantPhaseEstimation}, we will present a fault-tolerant version of phase estimation for approximately projecting the state in \cref{eq:SqueezedVacuumState} onto the $+1$ eigenspace of $S_p$ (see \cite{BarbaraDanielGKP} which provides the first description for using phase estimation to prepare approximate GKP states). 

\begin{figure} 
\centering
\includegraphics[height=7cm]{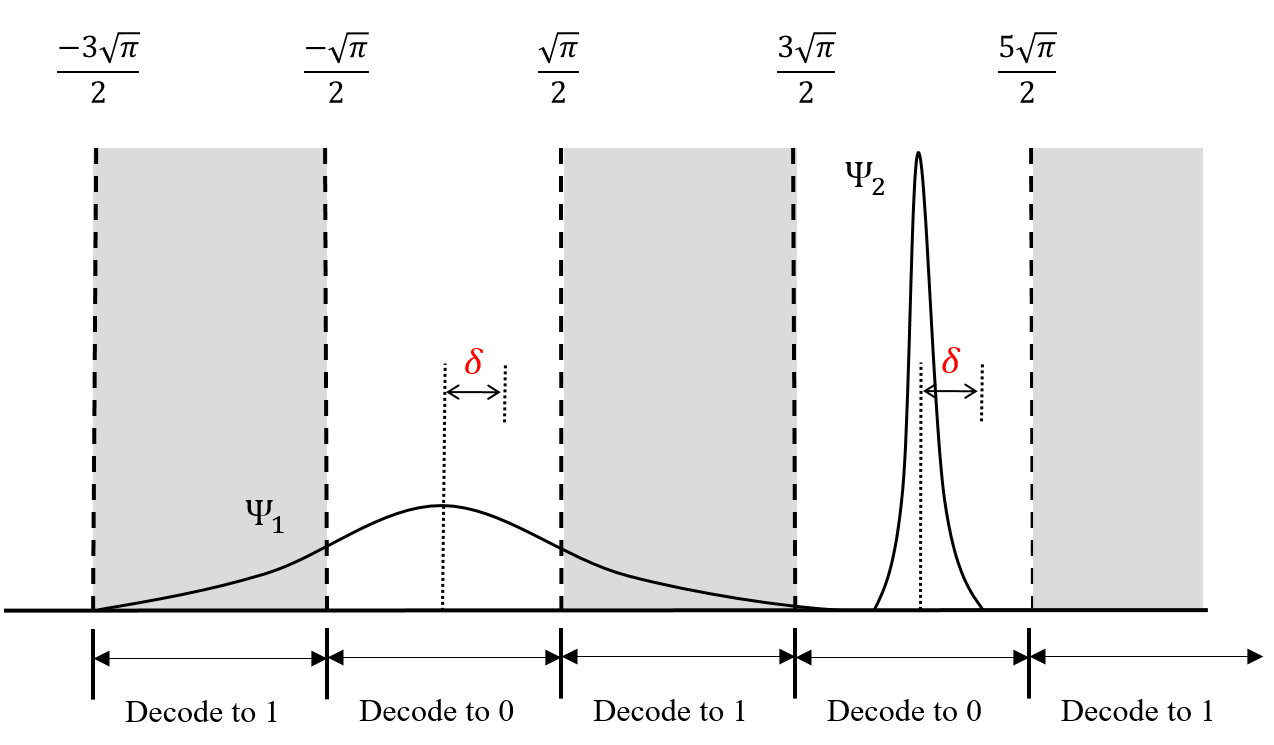}
\caption{Peaks centered at even integer multiples of $\sqrt{\pi}$ in $q$ space. The peak on the left contains large tails that extend into the region where a shift error is decoded to the logical $\ket{\overline{1}}$ state. The peak on the right is much narrower. Consequently for some interval $\delta$, the peak on the right will correct shift errors of size $\frac{\sqrt{\pi}}{2} - \delta$ with higher probability than the peak on the left.}
\label{fig:IllustrationPeaks}
\end{figure}

Naturally because of the finite width of the peaks of approximate GKP states, it will not be possible to correct a shift error in $p$ or $q$ of magnitude at most $\frac{\sqrt{\pi}}{2}$ with certainty. For example, suppose we have an approximate $\ket{\overline{0}}$ GKP state with a peak at $q=0$ subject to a shift error $e^{-ivp}$ with $|v| \le \frac{\sqrt{\pi}}{2}$. The finite width of the Gaussian peaks will have a non-zero overlap in the region $\frac{\sqrt{\pi}}{2} < q < \frac{3 \sqrt{\pi}}{2}$ and $\frac{-3 \sqrt{\pi}}{2} < q < \frac{-\sqrt{\pi}}{2}$. Thus with non-zero probability the state can be decoded to $\ket{\overline{1}}$ instead of $\ket{\overline{0}}$ (see \cref{fig:IllustrationPeaks} for an illustration). 

In general, if an approximate GKP state is afflicted by a correctable shift error, we would like the probability of decoding to the incorrect logical state to be as small as possible. A smaller overlap of the approximate GKP state in regions in $q$ and $p$ space that lead to decoding the state to the wrong logical state will lead to a higher probability of correcting a correctable shift error by a perfect GKP state. These remarks motivate the following definition
\begin{definition}
Let $\ket{\tilde{0}}$ be an approximate logical $\ket{\overline{0}}$ GKP state. We say that $\ket{\tilde{0}}$ is $(\frac{\sqrt{\pi}}{2} - \delta, \epsilon)_{q}$-GKP correctable if and only if for a given tuple $(\delta, \epsilon)$ with $\delta \le \frac{\sqrt{\pi}}{2}$, and $0 \le \epsilon \le 1$, we have
\begin{align}
\sum_{k = -\infty}^{\infty} \int_{2\sqrt{\pi}k - \delta}^{2\sqrt{\pi}k + \delta} | \langle q | \tilde{0} \rangle |^{2} dq > 1-\epsilon. 
\label{eq:Bound1}
\end{align}
We say that $\ket{\tilde{0}}$ is $(\frac{\sqrt{\pi}}{2} - \delta, \epsilon)_{p}$-GKP correctable if and only if for a given tuple $(\delta, \epsilon)$ with $\delta \le \frac{\sqrt{\pi}}{2}$, and $0 \le \epsilon \le 1$, we have
\begin{align}
\sum_{k = -\infty}^{\infty} \int_{\sqrt{\pi}k - \delta}^{\sqrt{\pi}k + \delta} | \langle p | \tilde{0} \rangle |^{2} dp > 1-\epsilon. 
\label{eq:Bound2}
\end{align}
\label{Def:DeltaEpsCorrect1}
\end{definition}
Note that the bounds in \cref{eq:Bound1,eq:Bound2} are different since GKP states have peaks defined on a rectangular lattice. Similarly, for an approximate logical $\ket{\overline{+}}$ state, we have the following definition
\begin{definition}
Let $\ket{\tilde{+}}$ be an approximate logical $\ket{\overline{+}}$ GKP state. We say that $\ket{\tilde{+}}$ is $(\frac{\sqrt{\pi}}{2} - \delta, \epsilon)_{p}$-GKP correctable if and only if for a given tuple $(\delta, \epsilon)$ with $\delta \le \frac{\sqrt{\pi}}{2}$, and $0 \le \epsilon \le 1$, we have
\begin{align}
\sum_{k = -\infty}^{\infty} \int_{2\sqrt{\pi}k - \delta}^{2\sqrt{\pi}k + \delta} | \langle p | \tilde{+} \rangle |^{2} dp > 1-\epsilon. 
\end{align}
We say that $\ket{\tilde{+}}$ is $(\frac{\sqrt{\pi}}{2} - \delta, \epsilon)_{q}$-GKP correctable if and only if for a given tuple $(\delta, \epsilon)$ with $\delta \le \frac{\sqrt{\pi}}{2}$, and $0 \le \epsilon \le 1$, we have
\begin{align}
\sum_{k = -\infty}^{\infty} \int_{\sqrt{\pi}k - \delta}^{\sqrt{\pi}k + \delta} | \langle q | \tilde{+} \rangle |^{2} dq > 1-\epsilon. 
\end{align}
\label{Def:DeltaEpsCorrect2}
\end{definition}

As an example, suppose we have two approximate GKP states $\ket{\tilde{0}}_{1}$ and $\ket{\tilde{0}}_{2}$ which are $(\frac{\sqrt{\pi}}{2} - \delta, \epsilon_{1})_{q}$ and $(\frac{\sqrt{\pi}}{2} - \delta, \epsilon_{2})_{q}$ correctable for a shift of size $e^{i(\frac{\sqrt{\pi}}{2} - \delta)p}$ (which is correctable by a perfect $\ket{\overline{0}}$ GKP state). If $\epsilon_{1} < \epsilon_{2}$, then $\ket{\tilde{0}}_{1}$ will correct a shift of size $\frac{\sqrt{\pi}}{2} - \delta$ with greater probability than $\ket{\tilde{0}}_{2}$. This is due to the fact that $\ket{\tilde{0}}_{1}$ has a smaller overlap in regions which result in decoding the logical $\ket{\overline{0}}$ state to the logical $\ket{\overline{1}}$. In this sense we say that $\ket{\tilde{0}}_1$ is better than $\ket{\tilde{0}}_2$ at correcting shift errors in $q$ space. 

\section{Fault-tolerant definitions}
\label{sec:FaultToltDefs}

Given a protocol for preparing an approximate GKP state, errors that occur during the protocol can potentially accumulate resulting in a large shift error in addition to deforming the output state. This can then result in an output state which significantly differs from a good approximate GKP state. By good, we mean a state that for a desired value of $\delta$, the state is $(\frac{\sqrt{\pi}}{2} - \delta, \epsilon_{1})_{q}$ and $(\frac{\sqrt{\pi}}{2} - \delta, \epsilon_{2})_{q}$ correctable with $\epsilon_1 , \epsilon_2 \ll 1$. 

The desired values for $\delta$ depend on the particular fault-tolerant error correction protocol used to correct shift errors on encoded data qubits (see for instance \cite{CDT09,AliferisCross07,CJL16,CJL16b,ChamberlandNeuralNet}). For example, one can use a version of the error correction scheme (which reduces to Steane error correction for qubit CSS stabilizer codes) as proposed by Glancy and Knill in \cite{GlancyKnillCondition}. In this scheme, logical $\ket{\overline{0}}$ and $\ket{\overline{+}}$ ancillas are prepared and interact with the encoded data qubit via a CNOT gate to correct the shift errors afflicting the data qubit. It is shown that the largest shift error that can be corrected is $\frac{\sqrt{\pi}}{6}$. The threshold of $\frac{\sqrt{\pi}}{6}$ arises from how shift errors that are initially afflicting the encoded data qubit and ancilla states propagate through the CNOT gates and combine prior to the measurement. In practice, if additional shift errors occur during the the error correction scheme (say after applying the CNOT gates), the largest correctable shift error could potentially be smaller than $\frac{\sqrt{\pi}}{6}$ \footnote{Using the optimizations considered in \cite{CDT09}, using Knill error correction could potentially increase the threshold of $\frac{\sqrt{\pi}}{6}$ to a larger value.}. In what follows, we will assume that after preparing the desired ancilla states using some state preparation protocol, the ancillas will be used in the error correction scheme of \cite{GlancyKnillCondition} (assumed to be fault-free) to correct shift errors on encoded data qubits. We also point out that due to the propagation of shift errors in the error correction scheme of \cite{GlancyKnillCondition}, it is important to prepare approximate $\ket{\tilde{0}}$ and $\ket{\tilde{+}}$ states that have small shift errors in \textit{both} $p$ and $q$.

If a small number of errors that occur during the preparation of an approximate GKP state result in large shift errors (or linear combinations of large shift errors) on the output state, then clearly the protocol used would not be practical. Thus it is important that a given state preparation protocol be fault-tolerant. In what follows we will define what we mean by fault-tolerant. We start with the following two definitions:
\begin{definition}
A shift error is said to be correctable if the magnitude of the shift is less than or equal to $\frac{\sqrt{\pi}}{6}$. Otherwise, we will say that the shift error is uncorrectable. 
\label{Def:CorrectableShiftErr}
\end{definition}
\begin{definition}
Suppose we have a protocol for preparing an approximate GKP state. We will say that the output state is an ideal approximate GKP state if no faults occur during the protocol.
\label{Def:IdealGKPstate}
\end{definition}
Thus by \cref{Def:IdealGKPstate}, any approximate GKP state obtained from a state preparation protocol will be called an ideal approximate GKP state if the protocol is implemented fault free, even if the output state is $(\frac{\sqrt{\pi}}{2} - \delta, \epsilon)_{p,q}$ correctable for some desired $\delta$ with large $\epsilon$ (so that the probability of correcting a shift $\frac{\sqrt{\pi}}{2} - \delta$ is small).

Note that the notion of correctable in \cref{Def:CorrectableShiftErr} assumes that only the data and ancilla qubits used in the error correction scheme of \cite{GlancyKnillCondition} have shift errors. If other operations such as the CNOT gates introduce additional errors, the correctable threshold would be smaller than $\sqrt{\pi}/6$. 

With the above definitions we are now ready to define what it means for a state preparation protocol of an approximate GKP state to be fault-tolerant. We point out that in \cref{Sec:FaultTolerantPhaseEstimation} we consider a fault-tolerant state preparation protocol based on phase estimation. Hence the definitions given below are specific to the case where an approximate GKP state is obtained by coupling a qubit to an oscillator.
\begin{definition}
\underline{$(m,\tilde{\delta})$-Fault-tolerant state preparation of an approximate GKP state}:

Suppose we have a protocol for preparing an approximate GKP state which is obtained by coupling qubits to a harmonic oscillator. Suppose also that at most $m$ faults occur during the protocol on the qubit Hilbert space and in addition, a correctable shift error in either $p$ or $q$ of size at most $\tilde{\delta}$ occurs on the oscillator Hilbert space. We will say that the protocol is an $(m,\tilde{\delta})$-Fault-tolerant state preparation of an approximate GKP state protocol if the output state differs from an ideal approximate GKP state by a correctable shift error. 
\label{Def:MfaultTolDef}
\end{definition}
A few clarifications are necessary. Firstly, a fault on the qubit Hilbert space corresponds to a location where an error can occur (see for instance \cite{AGP06}). By location we are referring to a particular time step where either a gate is implemented, a qubit is prepared, a qubit is measured or a qubit is idling. On the oscillator Hilbert space, if an error occurs, we can always expand that error into shift errors (see for instance Eq. (7.12) in \cite{AlbertBosonicCodePerformance18} and also \cite{BarbaraDanielGKP}). By performing the error correction scheme of \cite{GlancyKnillCondition}, measuring the $q$ and $p$ quadratures to perform error correction will always project the state onto a state with a single shift error in $q$ and $p$. Lastly, we point out that $\tilde{\delta}$ in \cref{Def:MfaultTolDef} relates to the largest allowed size of the shift error which occurs during the protocol. This should not be confused with $\delta$ in \cref{Def:DeltaEpsCorrect1,Def:DeltaEpsCorrect2} which relates to the size of a shift error that can be corrected by an approximate GKP state\footnote{The probability $1 - \epsilon$ of correcting a shift of size $\frac{\sqrt{\pi}}{2} - \delta$ depends on the location and width of the peaks.}.

Intuitively, a fault-tolerant protocol should have the property that if both the number of qubit errors and the size of shift errors are small, the resulting shift error on the output state should be correctable. Depending on the desired value for $\tilde{\delta}$, a particular fault-tolerant error correction protocol might be less tolerant to the size of the shift errors that occurred during the preparation of the approximate GKP state (in practice a different error correction scheme than Steane error correction could be used). 

Suppose a state preparation protocol satisfies \cref{Def:MfaultTolDef}. If during the preparation of the state, $m$ faults occur on the qubit Hilbert space in addition to a shift error of size at most $\tilde{\delta}$ on the oscillator Hilbert space, the remaining shift error on the output state will be corrected in a perfect version of the error correction protocol in \cite{GlancyKnillCondition}. As we will see in \cref{Sec:FaultTolerantPhaseEstimation}, due to measurement errors, the phase estimation protocol presented in \cite{BarbaraDanielGKP} needs to be modified in order to be a $(1,\tilde{\delta})$ fault-tolerant protocol. There are additional fault locations (apart from measurement errors) which can result in large shift errors that need to be treated with care. 

Lastly we describe how we will evaluate the error correction properties of an approximate GKP state obtained from a noisy state preparation protocol. It is important to compare the goodness of a GKP state prepared from a noisy circuit to that of an ideal circuit. If the output state of a noisy state preparation protocol is correctable, the shift error will be removed when performing error correction. Therefore we will compare the $(\frac{\sqrt{\pi}}{2} - \delta,\epsilon)_{p,q}$ correctable properties of output states after performing an optimal shift back correction (as long as the shift error is correctable) to the output state. If the shift error is not correctable, the protocol will be deemed too noisy. 

The optimal shift back correction is found as follows. In $q$ space, the optimal shift back $c^{q}_{\text{max}}$ is computed as
\begin{align}
    c_{\text{max}}^q=\argmax_{c}\sum_{k = -\infty}^{\infty} \int_{2\sqrt{\pi}k +c  - \delta}^{2\sqrt{\pi}k +c  + \delta} | \langle q | \tilde{0} \rangle |^{2} dq.
\end{align}
Similarly, in $p$ space we have 
\begin{align}
    c_{\text{max}}^p=\argmax_{c}\sum_{k = -\infty}^{\infty} \int_{\sqrt{\pi}k +c  - \delta}^{\sqrt{\pi}k +c  + \delta} | \langle p | \tilde{0} \rangle |^{2} dp.
\label{eq:ArgmaxShiftBack}
\end{align}

For protocols preparing $|\tilde{+}\rangle$, the metric can be defined analogously, but with the integral bounds for $p$ and $q$ switched.

\section{Fault-tolerant phase estimation protocol}
\label{Sec:FaultTolerantPhaseEstimation}

In \cref{subsec:FaulFreePhaseEstimation} we will give a brief review of the phase estimation protocol presented in \cite{BarbaraDanielGKP}. In \cref{subsec:FTPhaseEstOneDelt}, we will show how the protocol can be modified so that it becomes a $(1,\tilde{\delta})$-Fault-tolerant state preparation of an approximate GKP state and we will provide the value of $\tilde{\delta}$.

\subsection{Brief review of the phase estimation protocol for preparing approximate GKP states}
\label{subsec:FaulFreePhaseEstimation}

\begin{figure} 
\centering
\includegraphics[height=3.5cm]{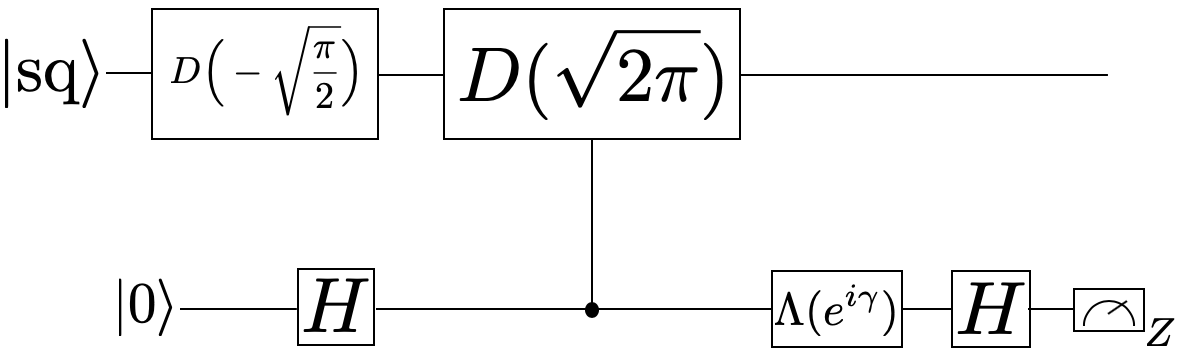}
\caption{Phase estimation circuit for preparing an approximate $\ket{\tilde{0}}$ GKP state. The initial state of the cavity is taken to be the squeezed vacuum state defined in \cref{eq:SqueezedVacuumState}. The circuit effectively projects the input squeezed vacuum state onto an approximate eigenstate of the $S_p$ operator while at the same time estimating the phase of the eigenvalue. The phase can be computed so that the output state can be shifted back to an approximate $+1$ eigenstate of $S_p$.}
\label{fig:PhaseEstimationCircuit}
\end{figure}

The phase estimation circuit for preparing an approximate $\ket{\tilde{0}}$ GKP state is given in \cref{fig:PhaseEstimationCircuit}. The $H$ gate is the Hadamard gate and $\Lambda(e^{i \gamma}) = \text{diag}(1,e^{i\gamma})$. After applying several rounds of the circuit in \cref{fig:PhaseEstimationCircuit}, the input squeezed vacuum state (given in \cref{eq:SqueezedVacuumState}) is projected onto an approximate eigenstate of $S_p$ with some random eigenvalue\footnote{The phase $\theta$ that is obtained depends on the measurement outcome of the ancilla qubit in each round.} $e^{i\theta}$. Additionally, an estimated value for the phase $\theta$ is obtained. After computing the phase, the state can be shifted back to an approximate $+1$ eigenstate of $S_p$. 

There are a variety of ways to choose the phases $\gamma$ at each round in the gate $\Lambda(e^{i \gamma})$. In a non-adaptive phase estimation protocol, half of the values for $\gamma$ can be chosen to be 0 and the other half can be chosen to be $\pi/2$. In an adaptive phase estimation protocol, if the phase to be estimated is $\theta$ and assuming no prior knowledge of $\theta$, during the first round the phase can be chosen as $\gamma_1 = 0$. For later rounds, it was shown in \cite{BarbaraDanielGKP} that the next phases can be chosen as
\begin{align}
\gamma_{m} = \argmax_{\gamma}{\sum_{x_{m} = 0,1}\Big | \int d\theta e^{i \theta} P_{\gamma}(x[m]| \theta)    \Big |}.
\label{eq:AdaptiveArgMax}
\end{align}
In \cref{eq:AdaptiveArgMax}, the probability $P_{\gamma}(x[m]| \theta)$ is the probability of obtaining measurement outcomes $x_1, \cdots  , x_m$ when the produced output eigenstate $\ket{\psi_{\theta}}$ has eigenvalue $e^{i \theta}$. Since the measurement results for each round are independent, the estimated probability is given by
\begin{align}
P_{\gamma = \gamma_{m}}(x[m]| \theta) = \prod_{i=1}^{m} \cos^{2}{\Big ( \frac{\theta + \gamma_{i}}{2} + x_{i}\frac{\pi}{2} \Big )}.
\label{eq:ProbOfEstPhase}
\end{align}
By analytically computing the evolution of the state, the exact probability of obtaining the outcomes $x[M]$ is derived in \cref{APP:ProbabilitySection}.

Given the final measurement record $x[M]$, the estimated phase $\tilde{\theta}$ is chosen as
\begin{align}
\tilde{\theta} = \text{arg} \int_{-\pi}^{\pi} d\theta e^{i\theta} P ( x[M] | \theta    ).
\label{eq:FinalEstimatedPhase}
\end{align}
The shift back correction based on the estimated phase $\tilde{\theta}$ is given by $e^{i\frac{\tilde{\theta}}{2\sqrt{\pi}}q}$. Note that for either the adaptive or non-adaptive protocol, the estimated phase and probabilities are computed using \cref{eq:ProbOfEstPhase,eq:FinalEstimatedPhase}.


\subsection{$(1,\tilde{\delta})$ fault-tolerant state preparation using phase estimation}
\label{subsec:FTPhaseEstOneDelt}

From the circuit of \cref{fig:PhaseEstimationCircuit} and from \cref{eq:FinalEstimatedPhase}, a measurement readout error can result in the wrong estimated phase which in turn results in an incorrect shift-back correction (application of $e^{ivq}$ where $v$ is computed using \cref{eq:AdaptiveArgMax,eq:ProbOfEstPhase}). Now suppose that the output state is afflicted by a shift error of the form $e^{-iwq}$. In this case, the final output state will have a total shift error of the form $e^{-i(v-w)q}$. If $|v-w| \mod \sqrt{\pi} > \sqrt{\pi}/6$, then the shift error will be uncorrectable and thus the phase estimation protocol will not be fault-tolerant \footnote{Note that $|v-w|$ should be taken modulo $\sqrt{\pi}$ since a perfect $\ket{\overline{0}}$ GKP state in $p$ space has peaks at any integer multiple of $\sqrt{\pi}$.}. In what follows we will identify an output state of the phase estimation protocol as $x[m]$ if it arises from the measurement outcomes $x_1 , \cdots, x_m$ in the fault-free case.

Consider the case where the phase estimation protocol is implemented in $m$ rounds and let us assume that a single measurement readout error occurs and that all other operations are fault free. In this case the output state $x_{\text{out}}[m]$ will have one bit which differs from the bit string $x_{\text{corr}}[m]$ that would have been obtained in the fault free case (so that the Hamming distance $d_{H}(x_{\text{out}}[m],x_{\text{corr}}[m]) = 1$). The shift-back operation of \cref{eq:AdaptiveArgMax} applied to the output state will be the shift correction of an output state that is one Hamming distance away from the actual state that was produced. Thus to ensure that the protocol is $(1,\tilde{\delta})$ fault-tolerant for some $\tilde{\delta}$, we should only accept output states $x_{\text{out}}[m]$ which have the property that applying the shift back correction corresponding to \textit{any} other state $x'[m]$ with $d_{H}(x'[m],x[m]) = 1$ results in a correctable left-over shift error. These remarks motivate the following protocol to prepare an approximate $\ket{\tilde{0}}$ state which is fault-tolerant to a single measurement readout error.

\vspace{25px}
\fbox{\begin{minipage}{45em}
        \textbf{Calculation of acceptance set $A_m$ and $\tilde{\delta}$.}

Consider all output states obtained from an $m$ round \textit{fault-free} phase estimation protocol of \cref{subsec:FaulFreePhaseEstimation}. Let $A_{m} = \emptyset$ (which we call the acceptance set) and $\Gamma_{m} = \emptyset$. For each output state $x[m]$, do the following

\begin{enumerate}
\item Compute the shift correction $v_{m} = \frac{1}{2\sqrt{\pi}}\text{arg} \int_{-\pi}^{\pi} d\theta e^{i\theta} P ( x[m] | \theta )$. 
\item Let $j[m]$ be a bit string of size $m$. For all $j \in \{ 1, \cdots , m \}$ such that $d_{H}(x[m], j[m]) = 1$, compute the shift correction $v^{(j)}_{m} = \frac{1}{2\sqrt{\pi}}\text{arg} \int_{-\pi}^{\pi} d\theta e^{i\theta} P ( j[m] | \theta )$. 
\item If $\nexists \thinspace \thinspace j$ such that $|v_{m} - v^{(j)}_{m}| \mod \sqrt{\pi} > \frac{\sqrt{\pi}}{6}$, append $x[m]$ to $A_{m}$ and $\forall \thinspace j \in \{ 1, \cdots, m \}$, append $|v_{m} - v^{(j)}_{m}|$ to $\Gamma_{m}$.

If $A_{m} \neq \emptyset$, $\tilde{\delta} = \max_{m} \Gamma_{m}$.

 \end{enumerate}
\end{minipage}}
\vspace{25px}

\fbox{\begin{minipage}{45em}
        \textbf{$(1,\delta)$ fault-tolerant $m$ round phase estimation protocol for measurement readout errors.}

Consider the acceptance set $A_{m} \neq \emptyset$ and $\tilde{\delta}$ computed from the procedure described above. During an implementation of the phase estimation protocol subject to measurement readout errors, if the obtained output state $x[m] \notin A_{m}$, abort the protocol and start anew.

\end{minipage}}
\vspace{25px}

Note that during the protocol, there can be more than one measurement readout error. However if more than one readout error occurs, it is possible that the output state is afflicted by a shift error of magnitude greater than $\sqrt{\pi}/6$. Our protocol only guarantees protection against a single readout error. 

\begin{table} [H]
	\begin{centering}
		\begin{tabular}{|c|c|c|c|c|c|}
			\hline
Four round phase estimation protocol acceptance set  $A_{4}$ & 1111 & 1010 & 0101 & 1000 & 0010 \\ \hline
Largest shift difference & 0.235 & 0.225 & 0.225 & 0.225 & 0.225 \\ \hline
Probability of obtaining output state & 0.1258 & 0.1287 & 0.1279 & 0.0505 & 0.0499 \\ \hline
		\end{tabular}
		\par\end{centering}		
	\caption{\label{Tab:AcceptMeasStringFourRoundsTable} The first row displays the accepted measurement strings (set $A_4$) for the four round non-adaptive phase estimation protocol. If the protocol does not output an element of $A_{4}$, the output is rejected and the protocol begins anew. The second row displays the largest shift difference that can occur when applying the phase estimated shift in the presence of a single measurement readout error. As can be seen, these are all less than $\sqrt{\pi}/6 \approx 0.295$. The third row gives the probabilities of obtaining each state at the output of the phase estimation protocol.}
\end{table}

As an example, consider the four round phase estimation protocol. Applying the procedure described above for the adaptive phase estimation protocol, we find that $A_{4}$ is empty when choosing the initial phase to be $\gamma_0 = 0$ or $\gamma_0 = \frac{\pi}{2}$. This indicates that the adaptive phase estimation protocol of four rounds described in \cref{subsec:FaulFreePhaseEstimation} is \textit{not} fault-tolerant to single measurement readout errors\footnote{However it is possible that the adaptive phase estimation protocol described in \cref{subsec:FaulFreePhaseEstimation} could be modified to be fault-tolerant to measurement readout errors. We leave such analysis to future work.}. If the phases are computed using the non-adaptive phase estimation protocol and applying the protocol described above, we find that $\tilde{\delta} = \frac{\sqrt{\pi}}{6} - 0.235 \approx 0.0604$. The set $A_{4}$ of accepted measurement strings is given in Table~\ref{Tab:AcceptMeasStringFourRoundsTable} and has $|A_{4}| = 5$. The total probability of obtaining a state in $A_{4}$ is roughly $48.3 \%$. For the phase estimation protocol with more than four rounds, the acceptance set and the acceptance probability is very sensitive to the initial phase $\gamma_0$ and the domain of the optimized $\gamma$. For example, if we choose the initial phase $\gamma_0$ to be $\frac{\pi}{2}$ and the range of $\gamma$ to be $[0,2\pi]$, we get 18 accepted states (for the adaptive protocol). In this case the total probability of obtaining a state in $A_{8}$ is $6.25\%$. If we choose the initial phase $\gamma_0$ to be 0, the acceptance set has four states and the probability of obtaining a state in $A_{8}$ is $1.3\%$. For the non-adaptive phase estimation protocol, $|A_{8}| = 66$ but the total probability of obtaining a state in $A_{8}$ is roughly $79.2\%$. These results indicate that although the adaptive phase estimation protocol outperforms the non-adaptive protocol in the fault-free case (\cite{BarbaraDanielGKP,DispSensorBarbara}), the adaptive phase estimation protocol cannot be used in the presence of  measurement readout errors for four rounds, and has significantly fewer states in $A_{8}$ for eight rounds. Therefore in a noisy implementation of phase estimation, the non-adaptive protocol is preferable. A list of the states belonging to the acceptance set $A_{8}$ for both the adaptive and non-adaptive protocols can be found at \url{https://github.com/godott/GKP_phase_estimation.git}.

Suppose now that for $m$ rounds the set $A_{m}$ is not empty. We then know there exists a $\tilde{\delta}$ such that the protocol is a $(1,\tilde{\delta})$ fault-tolerant $m$ round phase estimation protocol for \textit{measurement} readout errors. However there are other fault locations where a single fault resulting in a qubit error could potentially lead to an uncorrectable shift error on output states.  Note that an $X$ error prior to applying the controlled-$D(\sqrt{2 \pi})$ gate will do nothing since the qubit is in the $\ket{+}$ state. A $Z$ error will just change the sign of one of the peaks of the output state in $p$ space. However a damping event occurring on the ancilla qubit before or \textit{during} the application of the controlled-$D(\sqrt{2 \pi})$ gate can result in incorrectly applying the $D(\sqrt{2 \pi})$ displacement on the cavity. If several damping events occur during the protocol, the output state in $p$ space could potentially be badly deformed. In \cite{BarbaraDanielGKP} it was proposed to replace the ancilla qubit by a $k$-qubit cat state so that if a single qubit undergoes damping, a single shift error of size $D(\frac{\sqrt{2 \pi}}{k})$ would occur which can be made small for large $k$. However this approach would require the fault-tolerant preparation of a large $k$-qubit cat state which would significantly increase the required resources in addition to adding substantially more locations where errors can occur. Similar issues were observed in \cite{Rosenblum266} for preparing cat codes. However large cavity displacements were mitigated by interacting the cavity with a three level system $\ket{f}$, $\ket{e}$ and $\ket{g}$ instead of a qubit. A transition from $\ket{f}$ to $\ket{e}$ would not cause the incorrect gate from being applied. Thus two damping events would be required to cause the incorrect gate from being applied.

\begin{figure} 
\centering
\includegraphics[height=4.5cm]{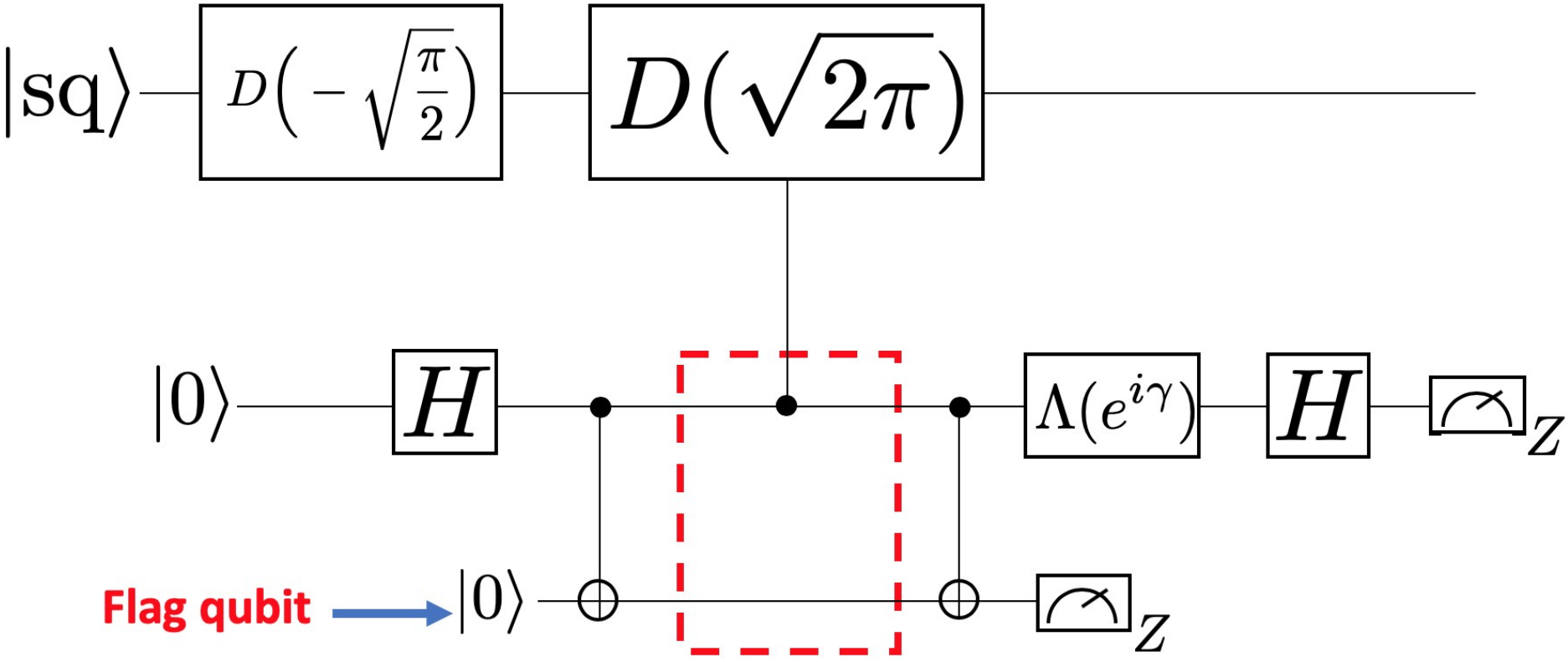}
\caption{Phase estimation circuit with an additional flag qubit. If a damping event occurs during the application of the controlled-$D(\sqrt{2 \pi})$ gate resulting in a $D(\sqrt{2 \pi})$ error, the flag qubit will be measured as 1 instead of 0. If the flag is measured as 1, we abort the protocol and start anew. }
\label{fig:OneFlagCircuitPhaseEstimation}
\end{figure}

Here we propose an alternative solution for mitigating damping errors during the application of the controlled-$D(\sqrt{2 \pi})$ gate which uses a single extra flag qubit \cite{CR17v1,CR17v2,CB17,TCD18Flag,ReichardtFlag18,CC18MagicFlag}. Consider the circuit in \cref{fig:OneFlagCircuitPhaseEstimation}. If a bit flip error occurs before or during the application of the controlled-$D(\sqrt{2 \pi})$ gate, the flag qubit will be measured as 1 instead of 0. In such a case the strategy will be simply to abort the protocol and start anew. The analysis for a general damping event using the flag qubit is given in \cref{sec:ControlDispAmpDamShiftErr}. It is shown that if ever a damping event on the $\ket{+}$ ancilla qubit causes the wrong $D(\sqrt{2 \pi})$ gate to be applied, the flag qubit will be measured as 1 instead of 0 in which case we abort the protocol and start anew. Thus with the flag qubit, the phase estimation protocol is robust to a damping error during the application of the controlled displacement gate. Note however that if a fault occurs in addition to a damping event, the shift error resulting from the damping event can potentially go undetected. For instance, if in addition to damping, the flag qubit is subject to a measurement error, the measurement outcome can be 0 instead 1 resulting in acceptance when the protocol should have been aborted.

Suppose that only a single damping event occurs during the four round phase estimation protocol, and that all other operations are fault-free. Using the analytic expressions derived in \cref{sec:ControlDispAmpDamShiftErr} and for a damping rate of $p=0.5$, we performed simulations to numerically characterize the effects of qubit damping on the prepared GKP state. We found that the shift error resulting from a single damping event was negligible compared to the largest shift error resulting from a single measurement readout error. 

Lastly, an $X$ error prior to the $\Lambda (e^{i \gamma})$ gate will change the sign of $\gamma$. After propagating through the Hadamard gate, the $X$ error will become a $Z$ error which will not affect the measurement outcome of the ancilla qubit. For the four round non-adaptive phase estimation protocol, we performed a simulation which showed that the shift error resulting from a single $X$ error prior to the $\Lambda (e^{i \gamma})$ gate is much smaller than the largest shift error arising from a measurement readout error on the ancilla qubit. 

Let us assume that the noise model afflicting the oscillator can cause a shift error of size at most $\tilde{\delta}$ in $p$ space (i.e. a shift of the form $e^{-i\tilde{\delta} q}$). From the above, the largest shift error that can arise from a single fault afflicting the qubit space during the four round fault-tolerant phase estimation protocol is $0.235< \sqrt{\pi}/6$. Following \cref{Def:MfaultTolDef}, we conclude that the protocol described in this section is a $(1,\tilde{\delta})$-fault-tolerant state preparation of an approximate logical $\ket{\overline{0}}$ GKP state with $\tilde{\delta} = \sqrt{\pi}/6 - 0.235 \approx 0.06$. 

In \cref{sec:CircuitQED} we will discuss a physical implementation of the phase estimation protocol using circuit QED. A much more detailed analysis of the noise afflicting the cavity and the resulting shift errors will be provided. 

\subsection{Goodness of approximate $\ket{\tilde{0}}$ states obtained from the noise free phase estimation protocol}
\label{sec:GoodnessNoiseFreeSection}

 \begin{figure} 
\centering
\begin{minipage}{.5\textwidth}
  \centering
  \includegraphics[width=0.9\linewidth]{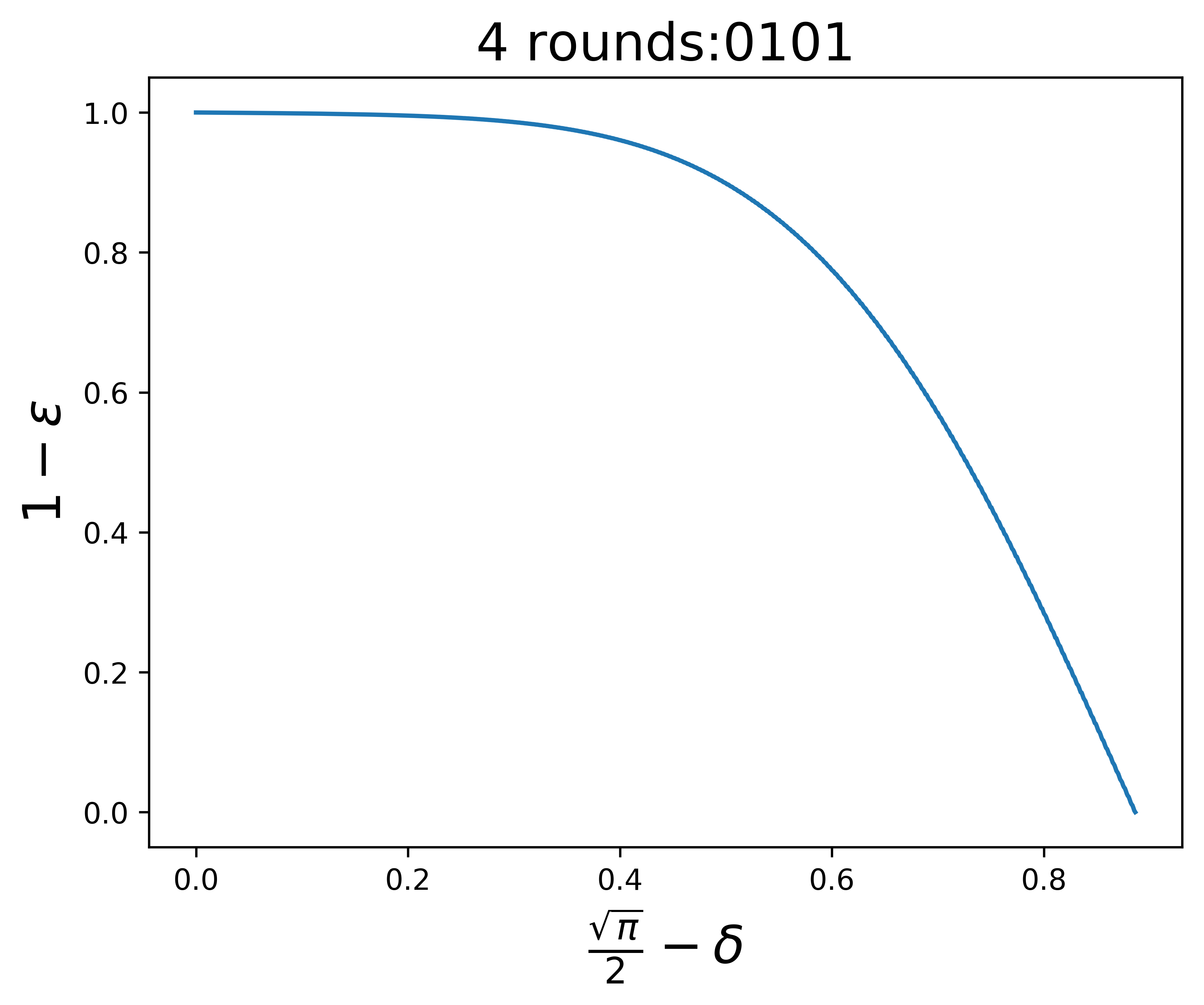}
  \label{fig:State0101EpsDeltaNoiseFree}
\end{minipage}%
\begin{minipage}{.5\textwidth}
  \centering
  \includegraphics[width=0.88\linewidth]{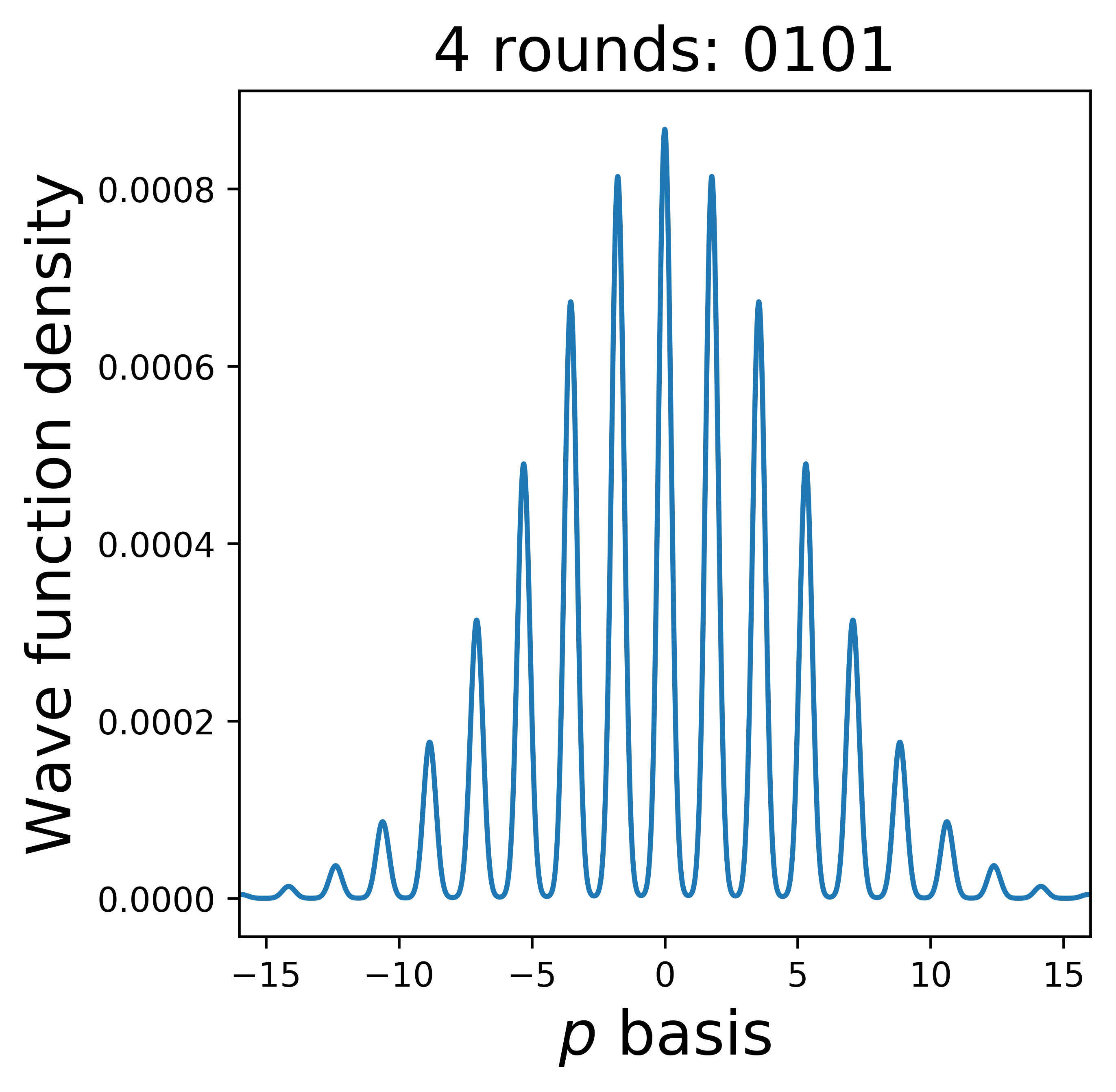}
  \label{fig:State0101PspaceNoiseFree}
\end{minipage}
\begin{minipage}{.5\textwidth}
  \centering
  \includegraphics[width=0.9\linewidth]{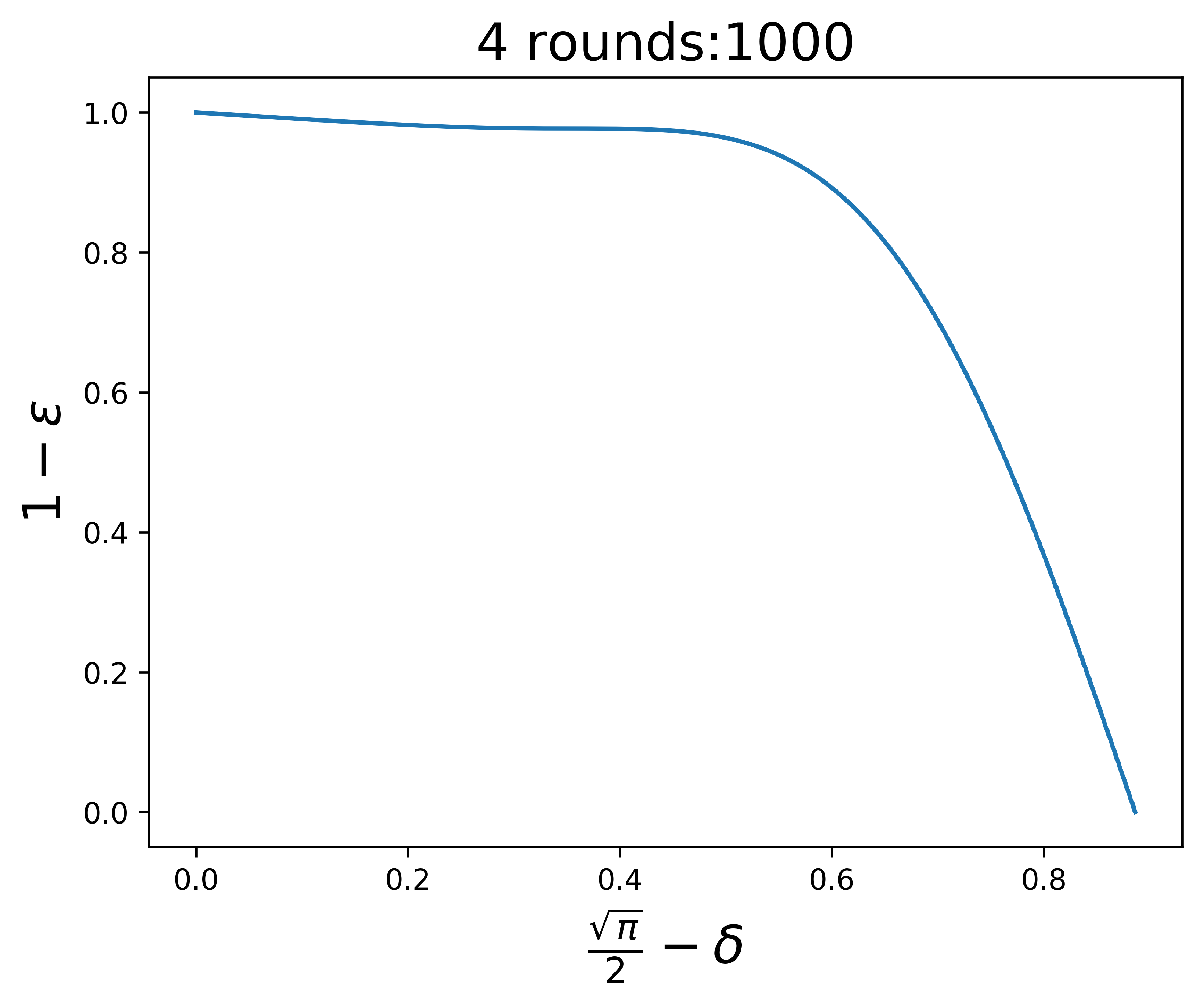}
  \label{fig:State1000EpsDeltaNoiseFree}
\end{minipage}%
\begin{minipage}{.5\textwidth}
  \centering
  \includegraphics[width=0.88\linewidth]{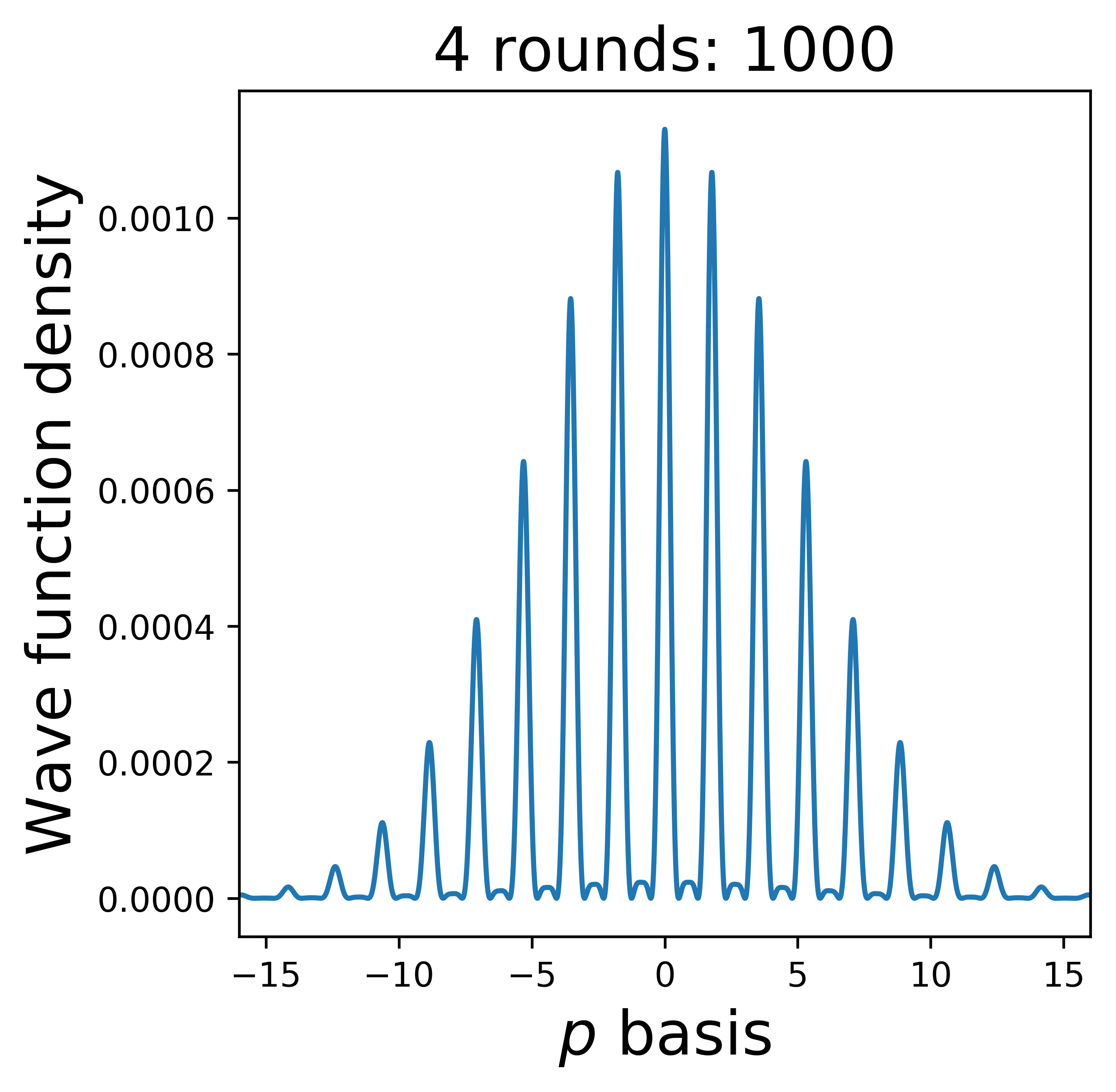}
  \label{fig:State1000PspaceNoiseFree}
\end{minipage}
\caption{(LEFT) Plot illustrating the probability of correcting a shift error of size $\frac{\sqrt{\pi}}{2} - \delta$ for the states 0101 and 1000 obtained via a non-adaptive noise free simulation of the phase estimation protocol presented in \cref{Sec:FaultTolerantPhaseEstimation}. (RIGHT) Wave function density $|\psi(p)|^2$ of the states 0101 and 1000 illustrated in $p$ space. The horizontal axis corresponds to values of $p$.}
\label{fig:EpsDeltaPlotsNoiseFree0101}
\end{figure}

In \cref{subsec:MasterEquationAnalysis}, the fault tolerant implementation of phase estimation described in this section is analyzed for a noise model which introduces gate noise, measurement readout errors and ancilla state preparation errors in addition to photon loss, amplitude damping and dephasing. Here we consider a noiseless implementation of the four round non-adaptive phase estimation protocol described in this section which will be used to benchmark the noisy implementation. 

Since the shift correction obtained from non-adaptive phase estimation is in $p$ space, for a range of values for $\delta$, we computed $\epsilon$ using the integral in \cref{Def:DeltaEpsCorrect1} after performing a shift back correction using \cref{eq:ArgmaxShiftBack}. This allowed us to compute the probability of correcting shift errors in $p$ of size $\frac{\sqrt{\pi}}{2} - \delta$. Plots for the states 0101 and 1000 which belong to $A_{4}$ (see \cref{Tab:AcceptMeasStringFourRoundsTable})  are given in \cref{fig:EpsDeltaPlotsNoiseFree0101}. The $(\frac{\sqrt{\pi}}{2} - \delta, \epsilon)_{p}$ correctable properties of the other states in $A_{4}$ are similar to the ones shown in \cref{fig:EpsDeltaPlotsNoiseFree0101}. It can be seen that for small values of size $\frac{\sqrt{\pi}}{2} - \delta$, both states can correct the shift error with probability close to 1. 

\section{Circuit QED implementation}
\label{sec:CircuitQED}

\subsection{State evolution in the dispersive regime}
\label{subsec:KerrDeriv}

In this section we will describe a direct implementation of the controlled-$D(\sqrt{2 \pi})$ gate. From \cite{BlaisFirstPaper,BlaisSecondPaper}, the Hamiltonian describing the coupling between a qubit and a cavity in the dispersive regime is given by

\begin{align}
H_{s} = \tilde{\omega}_{r} a^{\dagger}a + \tilde{\omega}_{a}Z + \chi a^{\dagger}aZ - \phi (a^{\dagger}a)^{2}Z,
\label{eq:SystemHamiltonianWithoutKerr}
\end{align}
where $\phi = \frac{g^4}{\Delta ^3}$, $\chi = \frac{g^2}{\Delta} - \phi$, $\tilde{\omega}_r = \omega_r + \phi$ and $\tilde{\omega}_a = \frac{\omega_a + \chi}{2}$. Here $g$ is the coupling strength between the qubit and the cavity and $\Delta = |\omega_r - \omega_a|$. The term $\phi (a^{\dagger}a)^{2}Z$ corresponds to the non-linear dispersive shift. The Hamiltonian in \cref{eq:SystemHamiltonianWithoutKerr} can be derived by performing an exact diagonalization of the Jaynes-Cummings Hamiltonian and keeping only leading order terms in $\phi$ \cite{BlaisSecondPaper}. Note that a more systematic treatment of the qubit as an anharmonic oscillator leads to an additional term in \cref{eq:SystemHamiltonianWithoutKerr} given by $-\frac{K}{2} (a^{\dagger}a)^2$ which is referred to as the Kerr non-linearity \cite{OptimalControl2}. Hence, in our analysis, we choose the system Hamiltonian to be given by
\begin{align}
H_{s} = \tilde{\omega}_{r} a^{\dagger}a + \tilde{\omega}_{a}Z + \chi a^{\dagger}aZ - \phi (a^{\dagger}a)^{2}Z - \frac{K}{2}(a^{\dagger}a)^{2}.
\label{eq:SystemHamiltonianWithKerr}
\end{align}

The direct implementation of the controlled-$D(\sqrt{2 \pi})$ gate can be achieved using the drive Hamiltonian
\begin{align}
H_{d}(t) = \mathcal{E}(t) a^{\dagger} e^{-i \omega_{d} t} + \mathcal{E}^{*}(t) a e^{i \omega_{d} t},
\end{align}
where $\mathcal{E}(t)$ describes the pulse shape of the drive and $\omega_d$ is the drive frequency. Thus the total Hamiltonian describing the evolution of the qubit-cavity system during the implementation of the controlled-$D(\sqrt{2 \pi})$ gate is given by
\begin{align}
H(t) = H_{s} + H_{d}(t). 
\label{eq:TimeDependentTotalH}
\end{align}

Going into the rotating frame of the qubit and the cavity and defining $\phi_{\pm} = \phi \pm K$ and $\omega_{\pm} = \pm \chi$, in \cref{App:NoiseAnalysisSection} we show that
\begin{align}
V_{R}(0,T) &= \mathcal{T} e^{-i \int_{0}^{T}H(t')dt'} \nonumber \\
&= R_{+}(T)D(A_{+})e^{iB_{+}} \ket{0}\bra{0} +
R_{-}(T)D(A_{-})e^{iB_{-}} \ket{1}\bra{1},
\label{eq:UnitaryRotatingFrameMain}
\end{align}
where

\begin{align}
R_{\pm}(T) = e^{-iT \omega_{\pm} a^{\dagger}a}(1 \pm iT \phi_{\pm} (a^{\dagger}a)^{2}),
\label{eq:RtermKerr}
\end{align}

\begin{align}
A_{\pm} = -i \int_{0}^{T} \mathcal{E}(t) e^{i \omega_{\pm} t}dt \mp \phi_{\pm}(2a^{\dagger}a-1)\int_{0}^{T} \mathcal{E}(t)te^{i\omega_{\pm}t}dt + \mathcal{O}(\phi_{\pm}^2),
\label{eq:APlusMinusTermKerr}
\end{align}

\begin{align}
B_{\pm} &= -\frac{i}{2}\int_{0}^{T}dt \int_{t}^{T}dt' \Big ( \mathcal{E}^{*}(t') \mathcal{E}(t) e^{i \omega_{\pm} (t' - t)} -   \mathcal{E}(t') \mathcal{E}^{*}(t) e^{-i \omega_{\pm} (t' - t)} \Big ) \nonumber \\  &\pm \frac{\phi_{\pm}(2a^{\dagger}a + 1)}{2}\int_{0}^{T}dt \int_{t}^{T}dt' \Big ( \mathcal{E}^{*}(t') \mathcal{E}(t) e^{i \omega_{\pm} (t' - t)} +  \mathcal{E}(t') \mathcal{E}^{*}(t) e^{i \omega_{\pm} (t' - t)} \Big )(t'-t) + \mathcal{O}(\phi_{\pm}^2).
\label{eq:BPlusMinusTermKerr}
\end{align}
In deriving \cref{eq:RtermKerr,eq:APlusMinusTermKerr,eq:BPlusMinusTermKerr}, we kept only leading order terms in $\phi_{\pm}$ since the Hamiltonian in \cref{eq:SystemHamiltonianWithKerr} neglects higher order terms in $\phi$ and $K$. We keep only leading order terms since with current experimental parameter values, $\phi_{\pm}$ is roughly three orders of magnitude smaller than $\chi$.

Firstly, notice that the terms $R_{\pm}(T)$ introduce a relative rotation between the $\ket{0}$ and $\ket{1}$ state of the qubit. Neglecting the terms proportional to $\phi_{\pm}$, it was shown in \cite{BarbaraDanielGKP} that by choosing the total interaction time to be $T = \pi / \chi$, the relative rotation between $\ket{0}$ and $\ket{1}$ can be set to one. However due to the presence of the non-linear dispersive shift and Kerr terms in \cref{eq:RtermKerr}, it is clear that the relative rotation cannot be completely removed. Further, $R_{\pm}(T)$ does not depend on the pulse shape $\mathcal{E}(t)$ of the drive term. Therefore even with an optimized pulse shape (see below), the effects from the non-linear dispersive shift and the Kerr will cause an undesired relative rotation between the qubit states. Regardless, we will still choose the interaction time $T$ to be $\pi / \chi$ to ensure that we eliminate the relative rotation from the leading order terms in \cref{eq:RtermKerr}. In \cref{subsec:KerrNonDisperUnitary}, we provide further details using the analytic expressions to show that to mitigate the effects due to the non-linear dispersive shift and Kerr terms would require the protocol to take place on time scales of order $\frac{1}{\phi_{\pm}}$ which (for the parameter values in \cref{Tab:ParameterValuesMasterEquation}) is four orders of magnitude larger than $\frac{\pi}{\chi}$. For such long time scales, noise terms such as photon loss, dephasing and damping would render the protocol impractical. 

The terms in \cref{eq:UnitaryRotatingFrameMain} that perform the desired controlled displacement gate are given by $A_{\pm}$ in \cref{eq:APlusMinusTermKerr}. The goal is to choose a pulse shape that implements the desired gate while at the same time minimizes the contributions from the non-linear dispersive shift and the Kerr term (terms proportional to $\phi_{\pm}$ in \cref{eq:APlusMinusTermKerr}). In order to be experimentally relevant, it is important to choose a pulse shape that is accessible to near term experiments using 2D and 3D cavities. We chose a Gaussian pulse with the following parameters
\begin{align}
\mathcal{E}(t) \approx -2.09562 \Big ( \frac{\sqrt{2\pi}}{T} \Big )e^{-\frac{(t-\mu)^{2}}{2\sigma^{2}}},
\label{eq:PulseShapeDriveNumeric}
\end{align}
where $\mu = \frac{\pi}{2\chi}$ and $\sigma = \frac{\pi}{10 \chi}$. With this Gaussian pulse we obtain
\begin{align}
A_{\pm} \approx \mp \frac{\sqrt{2 \pi}}{2} \mp \phi_{\pm}(2a^{\dagger}a-1)\int_{0}^{T} \mathcal{E}(t)te^{i\omega_{\pm}t}dt.
\label{eq:AplusMinChiIndependent}
\end{align}
The amplitude of the Gaussian, given by $ -2.09562 \Big ( \frac{\sqrt{2\pi}}{T} \Big )$, is chosen numerically to ensure that the appropriate gate is being performed. 

Since both the $A_{+}$ and $A_{-}$ terms are symmetric with opposite signs, the gate $D \Big( - \sqrt{\frac{\pi}{2}}  \Big )$ in \cref{fig:OneFlagCircuitPhaseEstimation} is \textit{not} required. Furthermore, the first term in \cref{eq:AplusMinChiIndependent} is independent of $\chi$. However the contribution from the second term in \cref{eq:AplusMinChiIndependent} will depend on the coupling strength and relative frequencies between the qubit and the cavity. Thus reducing the value of $\chi$ will minimize the undesired effects arising from the non-linear dispersive shift and the Kerr term while still producing the desired gate.

The $B_{\pm}$ terms of \cref{eq:BPlusMinusTermKerr} result in a phase difference between the $\ket{0}$ and $\ket{1}$ qubit states. However, computing the first integrals in \cref{eq:BPlusMinusTermKerr} (i.e. terms before $\phi_{\pm}$), we find that for our chosen pulse, the phase difference remains \textit{fixed} during every round of the phase estimation protocol. Therefore after applying the controlled displacement gate, we apply an additional phase gate which removes the phases introduced by the left most integrals of $B_{\pm}$. Note however that the $\phi_{\pm}$ terms in $B_{\pm}$ will not be cancelled and will thus introduce additional shift errors.

\subsection{Full noise analysis and master equation results.}
\label{subsec:MasterEquationAnalysis}

 In this section we perform a numerical analysis for the noisy implementation of the phase estimation protocol described in \cref{subsec:FTPhaseEstOneDelt}. The simulation is performed in several steps that we now describe. First, the controlled displacement gate is modeled using the following master equation
\begin{align}
\dot{\rho} = -i [H(t),\rho] + \kappa \mathcal{D}[a]\rho + \gamma_{1} \mathcal{D}[\sigma_{-}]\rho + \gamma_{\phi} \mathcal{D}[\sigma_{z}]\rho,
\label{eq:MasterEquationWithLindblad}
\end{align}
where $H(t)$ is given by \cref{eq:TimeDependentTotalH} after going into the rotating frame and $\mathcal{D}[L]\rho = (2 L \rho L^{\dagger} - L^{\dagger}L\rho - \rho L^{\dagger}L)/2$. The density matrix corresponds to the joint state of the cavity, ancilla qubit and flag qubit. The parameters $\kappa$, $\gamma_{1}$ and $\gamma_{\phi}$ correspond to the photon loss rate, the qubit decay rate and qubit dephasing. The pulse shape of the drive is given by \cref{eq:PulseShapeDriveNumeric}. The total evolution time of the controlled displacement is given by $\frac{\pi}{\chi}$ with $\chi = \frac{g^2}{\Delta} - \phi$.  

\begin{table} 
	\begin{centering}
		\begin{tabular}{|c|c|c|}
			\hline
Parameter values & Simulation 1 & Simulation 2 \\ \hline
$p$ & $5 \times 10^{-3}$ & $10^{-3}$  \\ \hline
$\frac{\kappa}{2\pi}$ & $1.59 \times 10^{-5}$MHz & $1.59 \times 10^{-6}$MHz   \\ \hline
$\frac{\gamma_{1}}{2\pi}$ & $1.06 \times 10^{-3}$MHz & $1.06 \times 10^{-4}$MHz   \\ \hline
$\frac{\gamma_{\phi}}{2\pi}$ & $7.96 \times 10^{-4}$MHz & $7.96 \times 10^{-5}$MHz   \\ \hline
$g$ & $8.92$MHz & $5.09$MHz   \\ \hline
$\Delta$ & $0.32$GHz & $0.32$GHz   \\ \hline
$K$ (Kerr) & $10^{-4}$MHz & $10^{-5}$MHz   \\ \hline
		\end{tabular}
		\par\end{centering}		
	\caption{\label{Tab:ParameterValuesMasterEquation} Parameter values for the two simulations of the non-adaptive noisy phase estimation protocol. The values for $\kappa$ are chosen based on state of the art 3D cavities. The parameters in the second column results in a $T_1$ time given by $T_1 = \frac{2\pi}{\kappa} = 10\text{ms}$. For a resonator frequency $f_r = 7\text{GHz}$, the quality factor is given by $Q = \frac{f_{r}}{\kappa/(2\pi)} = 7 \times 10^7$. For all simulations, the squeezing level of the input squeezed vacuum state is chosen to be 0.2. }
\end{table}

\begin{figure} 
\centering
\begin{minipage}{.5\textwidth}
  \centering
  \includegraphics[width=0.98\linewidth]{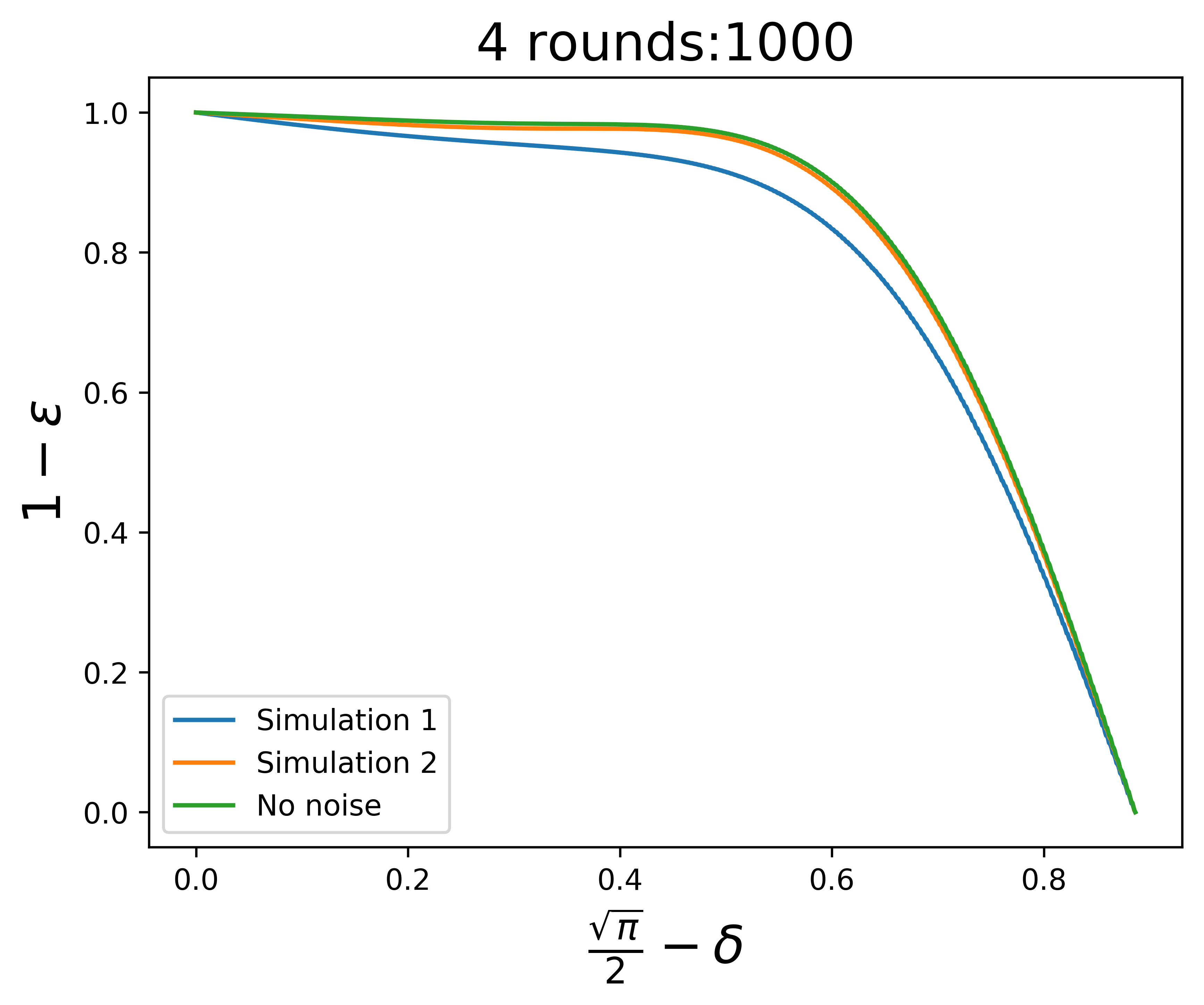}
  \label{fig:Kerr10State1000}
\end{minipage}%
\begin{minipage}{.5\textwidth}
  \centering
  \includegraphics[width=0.98\linewidth]{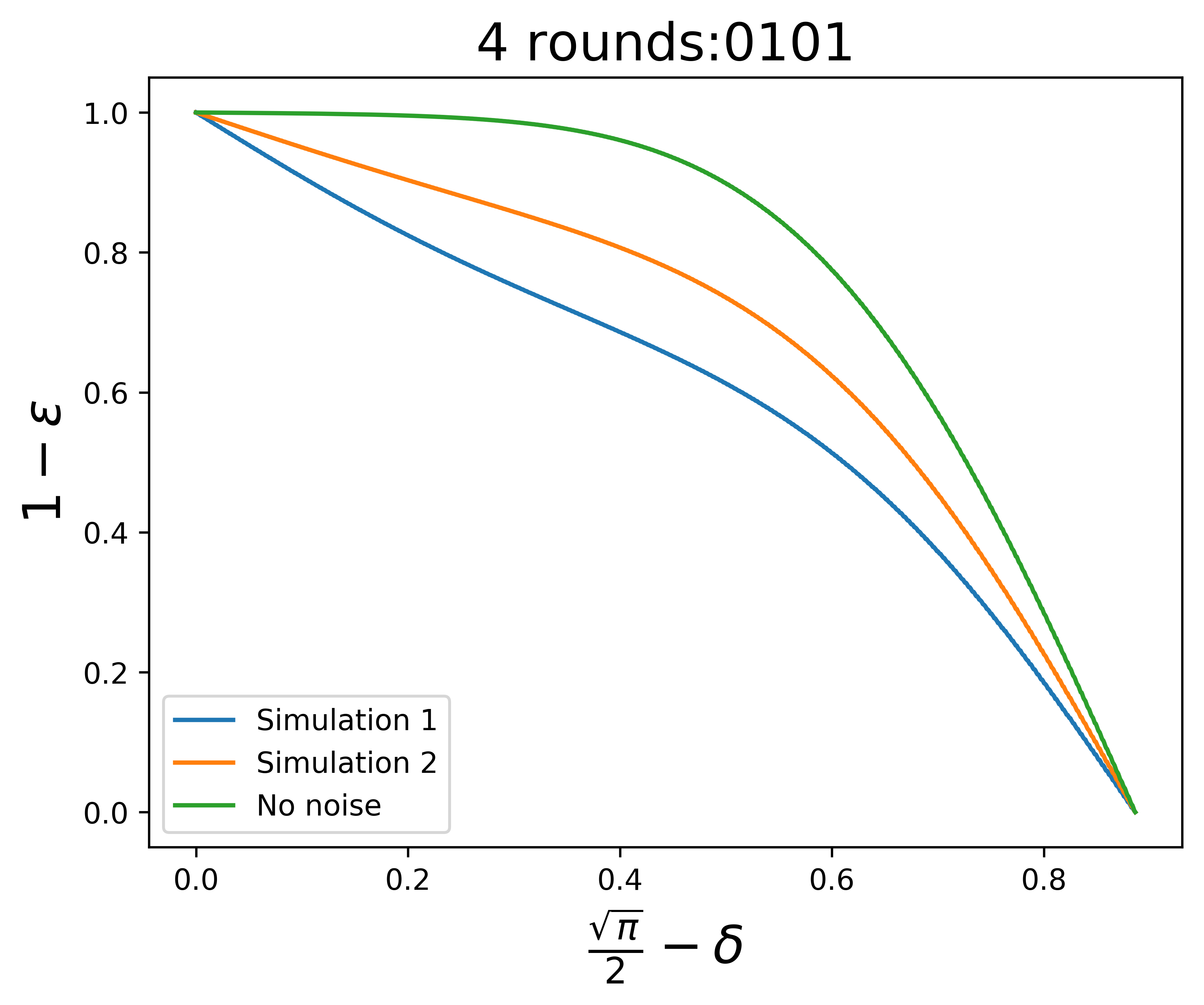}
  \label{fig:Kerr100State0101}
\end{minipage}
\begin{minipage}{.5\textwidth}
  \centering
  \includegraphics[width=0.98\linewidth]{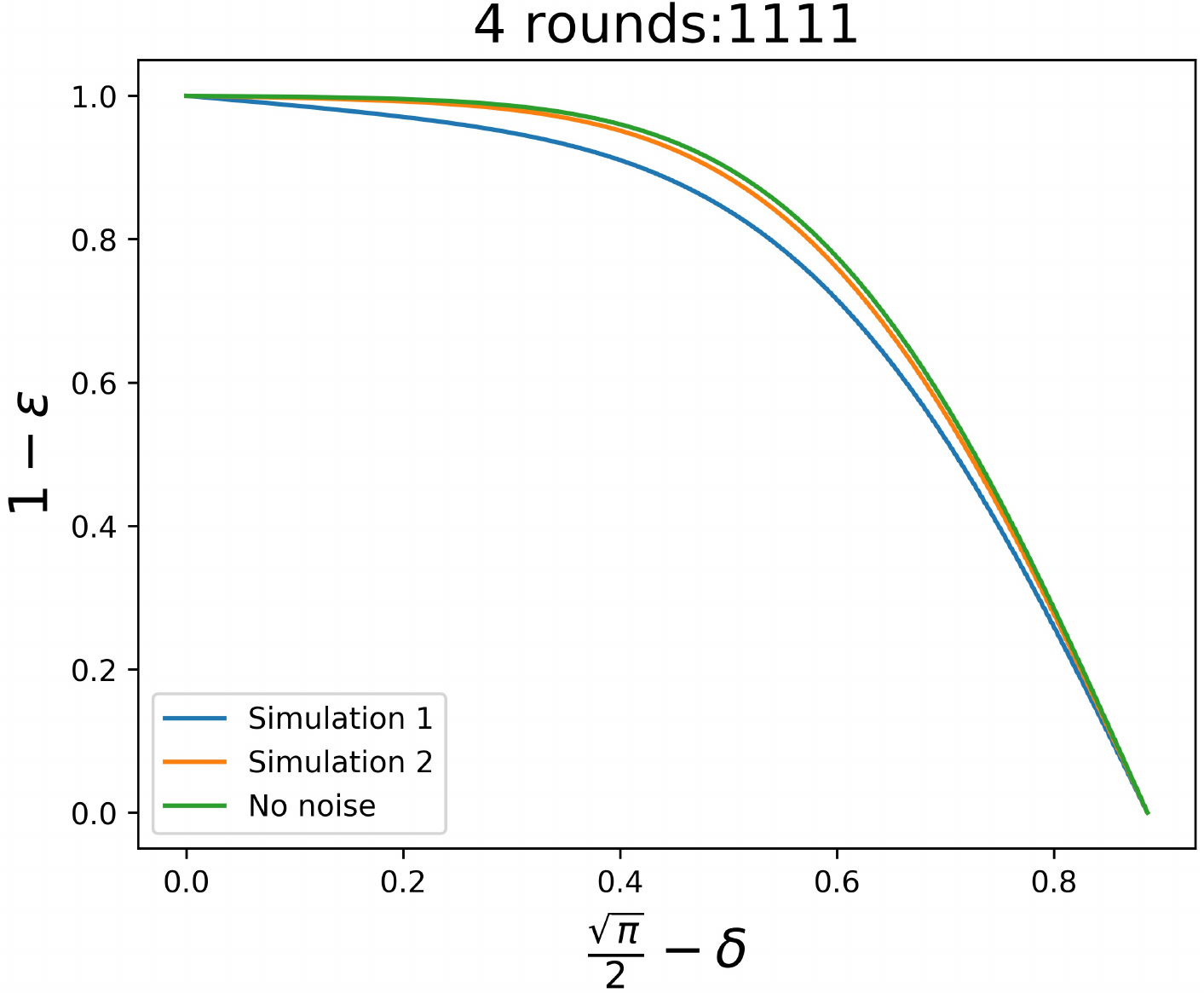}
  \label{fig:Kerr100State1111}
\end{minipage}
\caption{Probability of correcting a shift errors of size $\frac{\sqrt{\pi}}{2} - \delta$ for parameters chosen from the second (labeled "Simulation 1") and third (labeled "Simulation 2") column of \cref{Tab:ParameterValuesMasterEquation} in addition to the case where no noise is present. The plots are for the states 1000, 0101 and 1111 obtained from the four round fault-tolerant phase estimation protocol described in \cref{Sec:FaultTolerantPhaseEstimation}.}
\label{fig:Ker10AndKer100State0101}
\end{figure}

Both before and after the controlled displacement gate, the state of the cavity is subject to photon loss and its evolution is computed by solving a master equation. Based on current gate, state preparation and measurement times, we chose the evolution time from the preparation of the ancilla to the first CNOT gate (after which the controlled-displacement gate is performed) to be $0.14\mu$s. Similarly, the evolution time after the controlled-displacement gate to the time the ancilla is measured is also chosen to be $0.14\mu$s. Thus during one round, the cavity freely evolves with photon loss for a total of $0.28\mu$s. In addition, before and after the controlled displacement gate, we allow all qubit locations to fail with the following depolarizing noise model
\begin{enumerate}
	\item With probability $p$, each two-qubit gate is followed by a two-qubit Pauli error drawn uniformly and independently from $\{I,X,Y,Z\}^{\otimes 2}\setminus \{I\otimes I\}$.
	\item With probability $\frac{2p}{3}$, the preparation of the $\ket{0}$ state is replaced by $\ket{1}=X\ket{0}$. 
	\item With probability $\frac{2p}{3}$, any single qubit measurement has its outcome flipped. 
	\item With probability $p/10$, each Hadamard gate is followed by a Pauli error drawn uniformly and independently from $\{ X,Y,Z \}$.
\end{enumerate}
We chose $p/10$ for Hadamard gate failures since for current superconducting architectures, single qubit gate fidelities are about an order of magnitude higher than two-qubit gate fidelities. In practice, the gate errors applied to the qubit Hilbert space will depend strongly on the circuit QED architecture and should also be modeled using a master equation as in \cref{eq:MasterEquationWithLindblad}. However, we chose a depolarizing model in order to reduce the computation time and simplify the analysis. We also mention that in our analysis we assumed that $g$ is tunable. Thus both before and after the controlled-displacement gate, the cavity and qubit system is decoupled and can be treated separately. 

\begin{figure} 
\centering
\begin{minipage}{.5\textwidth}
  \centering
  \includegraphics[width=0.98\linewidth]{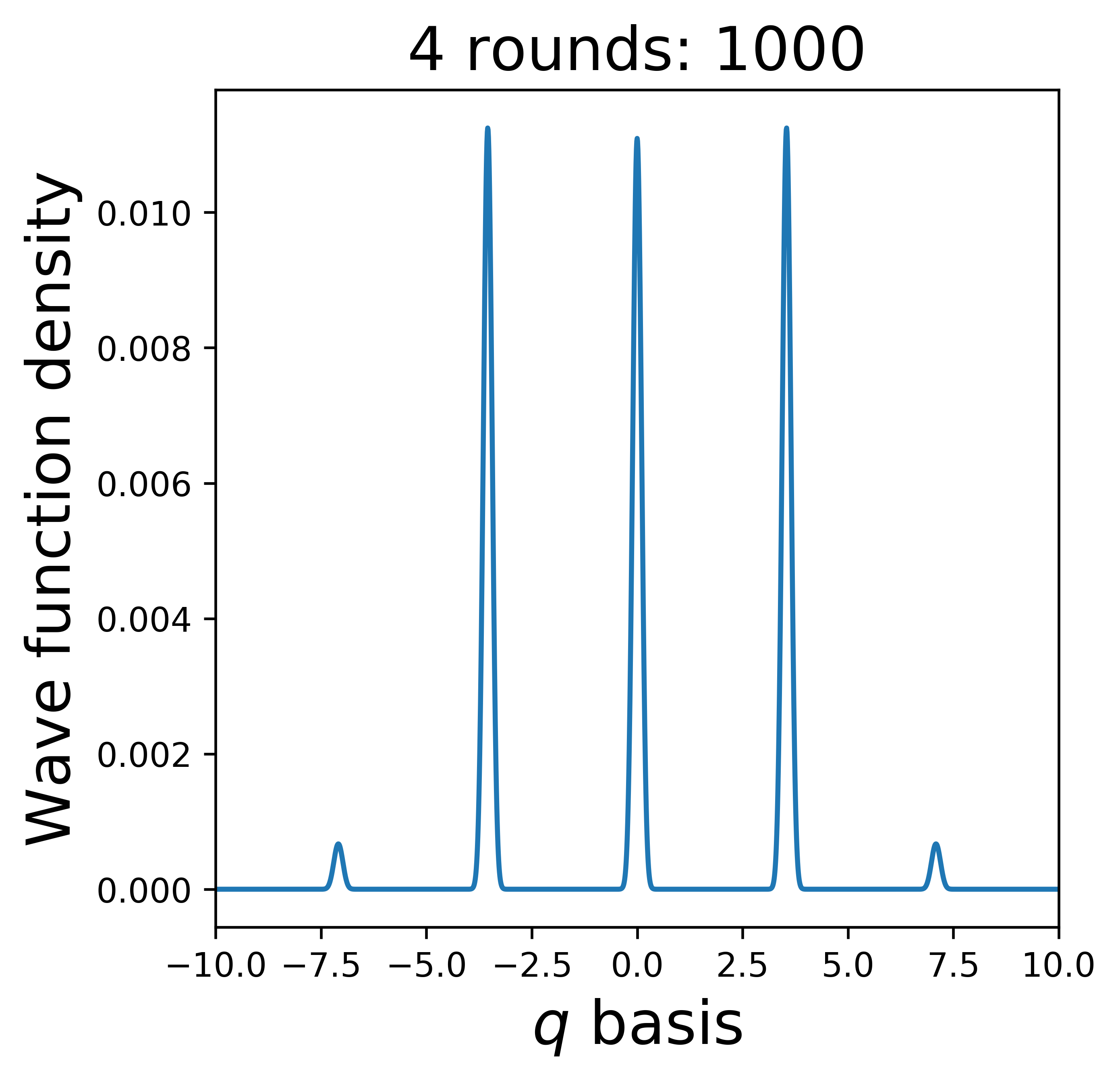}
  \label{fig:qbasis1000}
\end{minipage}%
\begin{minipage}{.5\textwidth}
  \centering
  \includegraphics[width=0.98\linewidth]{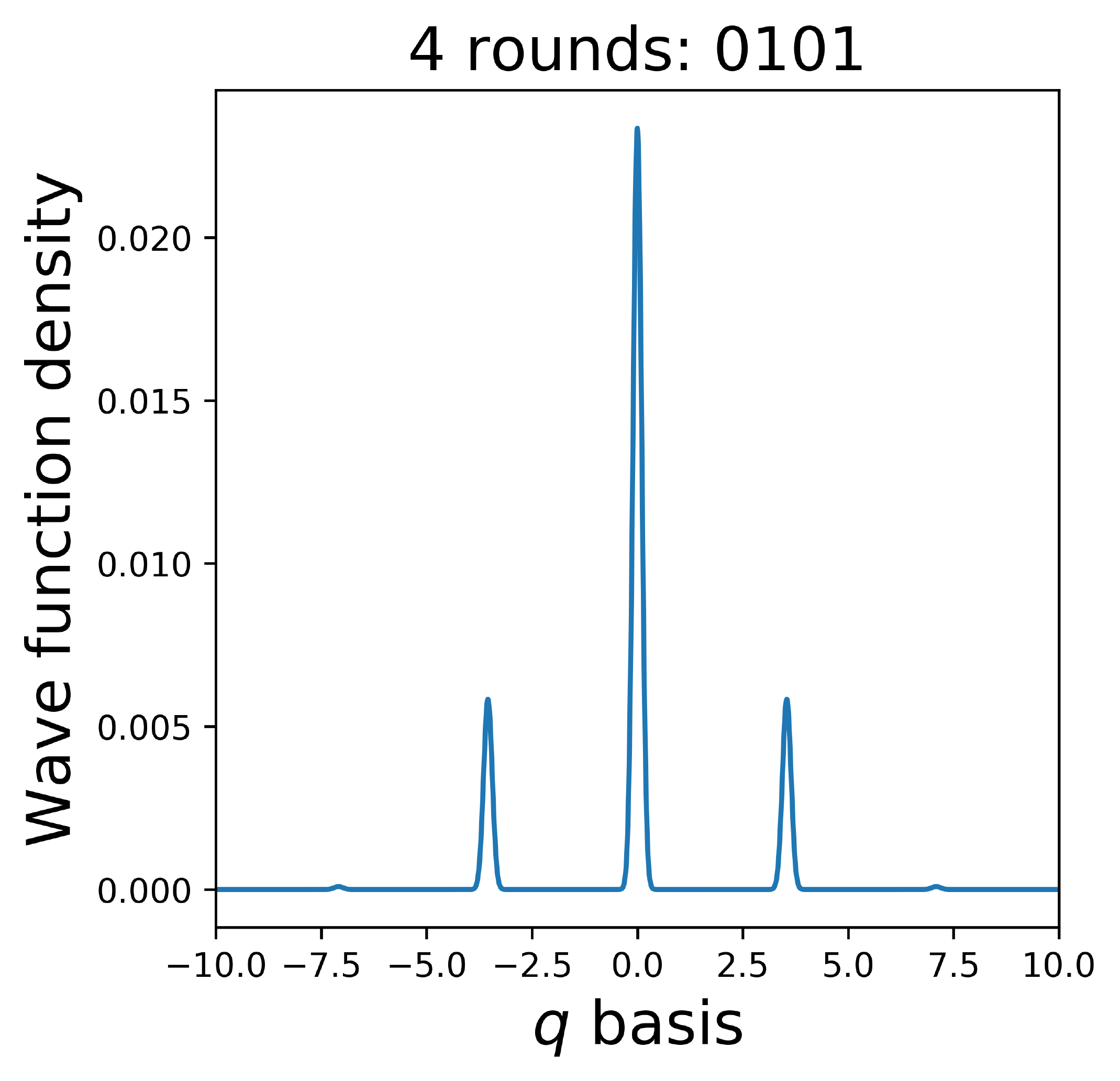}
  \label{fig:qbasis0101}
\end{minipage}
\begin{minipage}{.5\textwidth}
  \centering
  \includegraphics[width=0.98\linewidth]{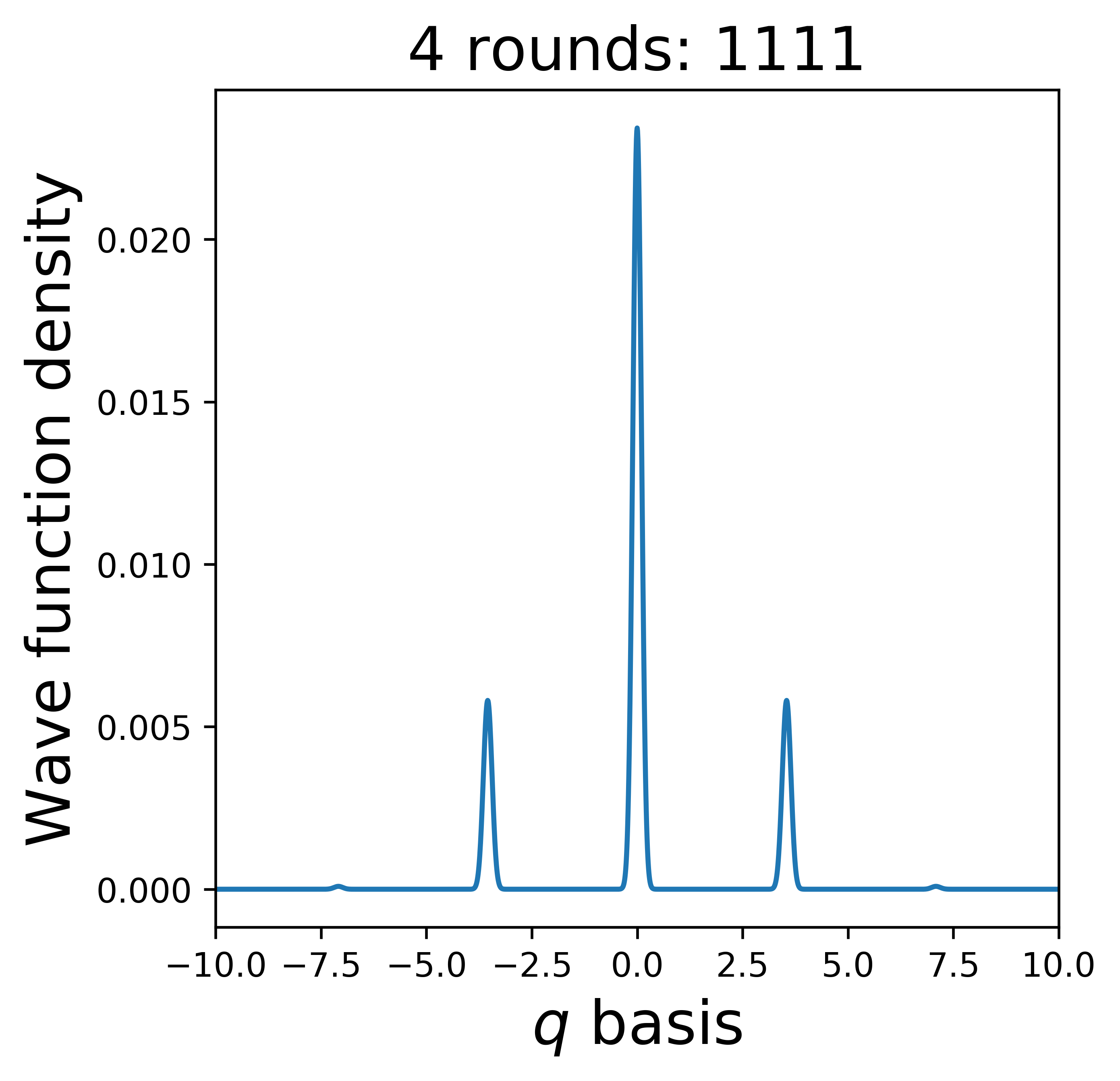}
  \label{fig:qbasis1111}
\end{minipage}
\caption{Wave function density $|\psi(q)|^2$ of the states 1000, 0101 and 1111 obtained from the noise free non-adaptive phase estimation protocol.}
\label{fig:qBasisGoodAndBadStates}
\end{figure}

The master equation was numerically solved using Qutip \cite{JOHANSSON20131234}, our code can be accessed at \url{https://github.com/godott/GKP_phase_estimation.git}. Due to the long computation time during the controlled displacement gate, performing a full Monte Carlo simulation to take into account gate errors with the depolarizing model was unfeasible (i.e. gate locations both before and after the controlled displacement gate). Instead, we analytically computed the error probabilities for each Pauli operator (using the depolarizing noise model described above) immediately after the first CNOT gate, before both measurement locations and before the phase gate\footnote{The error probabilities were computed by considering all possible single-fault locations resulting in a given error at the considered location. For instance, a $Z \otimes I$ error after the first CNOT gate can arise from a $Z \otimes I$ error from the faulty CNOT gate, but also from a $Z$ error after the application of the Hadamard gate prior to the CNOT gate.}.  At each location, all possible Pauli operators based on their associated probabilities were added. For a given Pauli error, the probabilities (which are expressed as functions of $p$) were then used (in addition to the state evolution obtained from the master equation) to compute the final probability of obtaining a given output state. Instances with two or more faults occurring on the qubit Hilbert space before and after the controlled-displacement gate were neglected. However our analysis is still more complete than previous implementations which considered only measurement errors and errors during the controlled-displacement gate. More details of the Pauli simulation can be found in \cref{APP:ProbabilitySection}.

\begin{table} 
	\begin{centering}
		\begin{tabular}{|c|c|c|c|c|c|}
			\hline
Four round phase estimation protocol acceptance set  $A_{4}$ & 1111 & 1010 & 0101 & 1000 & 0010 \\ \hline
Probability of obtaining output state noisy simulation 1 & 0.1297 & 0.0727 & 0.0737 & 0.0474 & 0.0429 \\ \hline
Probability of obtaining output state noisy simulation 2 & 0.1468 & 0.0982 & 0.0978 & 0.0514 & 0.0448 \\ \hline
		\end{tabular}
		\par\end{centering}		
	\caption{\label{Tab:ProbStatesFourRoundNoisy} The second row corresponds to the probabilities of obtaining the output states of $A_4$ for the parameters chosen from the second column of \cref{Tab:ParameterValuesMasterEquation} conditioned on all flag measurements being 0. The third row is identical but for the parameters chosen from the third column of \cref{Tab:ParameterValuesMasterEquation}. The probabilities are smaller for noisier circuits since the flag qubit has a higher chance of being measured as 1 causing the protocol to be aborted. }
\end{table}

We performed two different simulations where for each simulation, the chosen parameter values are given in \cref{Tab:ParameterValuesMasterEquation}. The parameters chosen for the first simulation (middle column in \cref{Tab:ParameterValuesMasterEquation}) are based on current experimental values for 2D and 3D cavities \cite{GKPparams1,Rosenblum266,GKPparams2,LongCavityLifetimes,2DCavityLifetimes,2DCavityLifetimesv2}. The parameters chosen for the second simulation are based on values that might be obtained with improved future technologies. Plots showing the probability of correcting shift errors $\frac{\sqrt{\pi}}{2} - \delta$ after performing a shift correction of the output states of the noisy phase estimation protocol using \cref{eq:ArgmaxShiftBack} are given in \cref{fig:Ker10AndKer100State0101}. In what follows we will refer to these plots as $\epsilon - \delta$ plots. Note that noise during the phase estimation protocol introduces more shift errors in $p$ space. Therefore in \cref{fig:Ker10AndKer100State0101}, only $\epsilon - \delta$ plots in $p$ space are shown. 

\begin{figure} 
\centering
\begin{minipage}{.5\textwidth}
  \centering
  \includegraphics[width=0.98\linewidth]{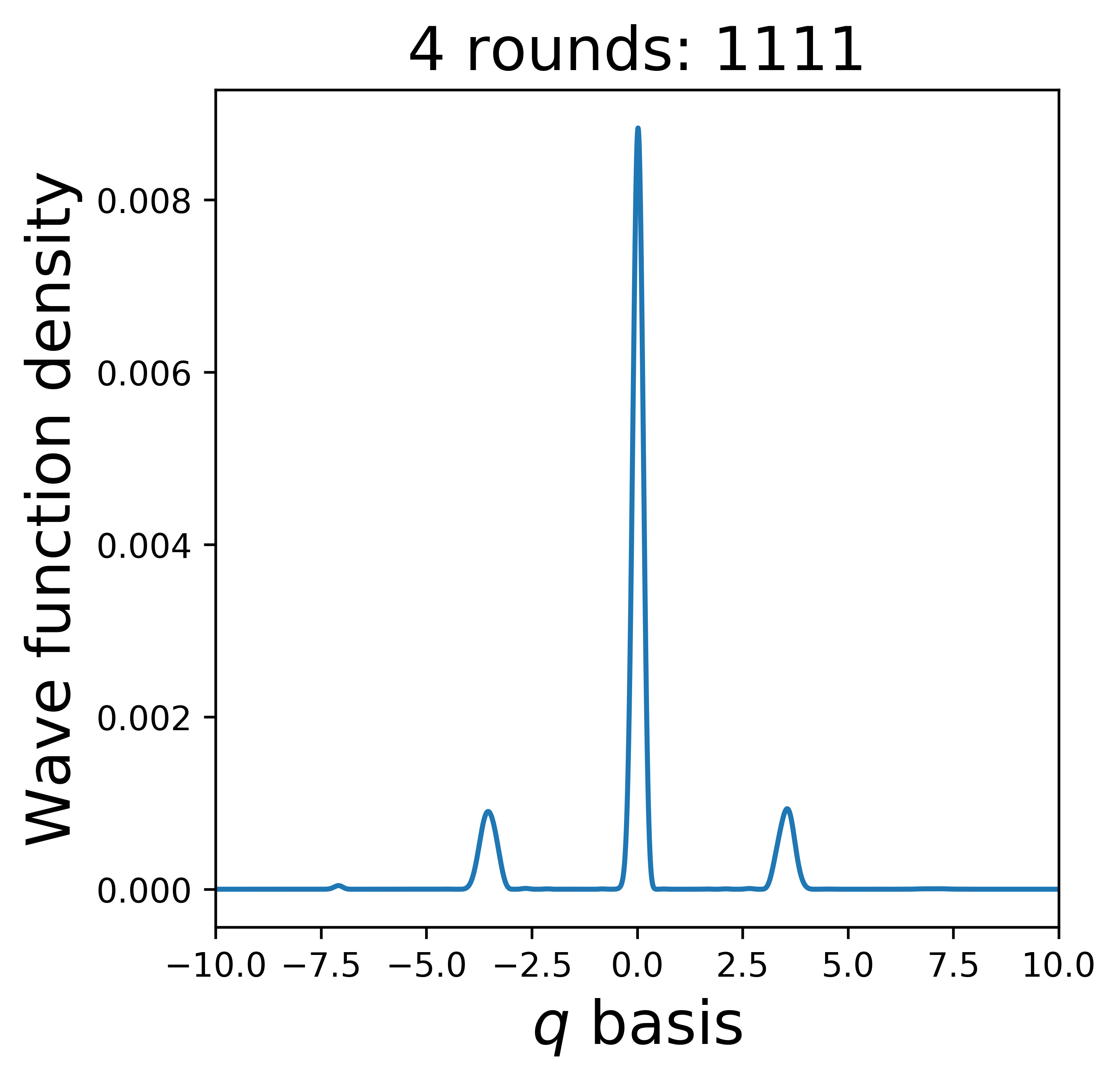}
  \label{fig:qbasis1111KerrOnly}
\end{minipage}%
\begin{minipage}{.5\textwidth}
  \centering
  \includegraphics[width=0.98\linewidth]{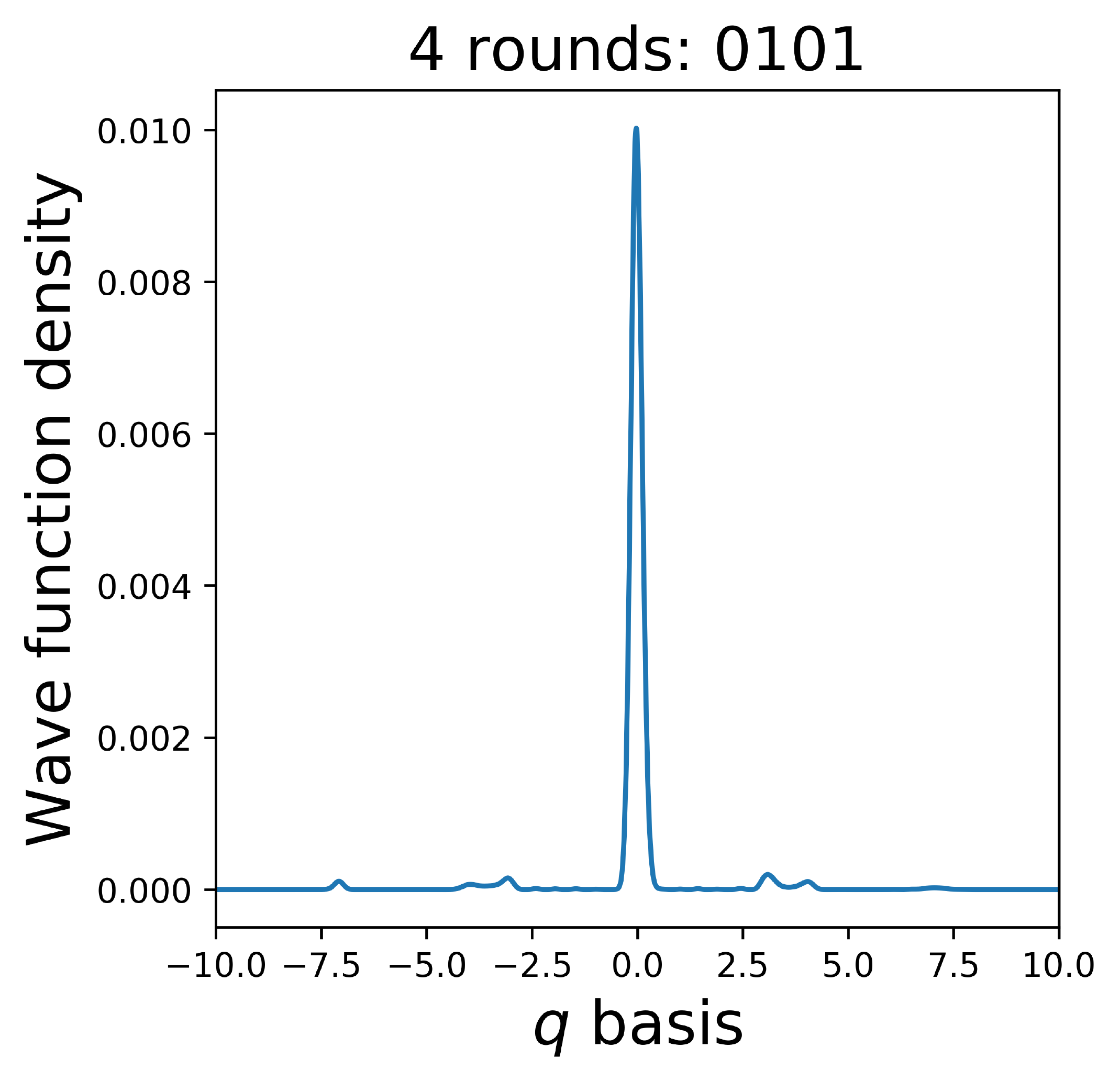}
  \label{fig:qbasis0101KerrOnly}
\end{minipage}
\caption{Wave function density of the states 1111 and 0101 where all noise terms are excluded apart from the Kerr and non-linear dispersive shift terms. Comparing to \cref{fig:qBasisGoodAndBadStates}, the Kerr and non-linear dispersive shift terms reduce the amplitudes of the two small peaks of the 0101 state significantly more than those for the 1111 state.}
\label{fig:KerrOnlyStateComparison}
\end{figure}

For the four round phase estimation protocol, we find that the state 1010 has a similar $\epsilon - \delta$ plot to the one for the state 0101 whereas the state 0010 has a similar $\epsilon - \delta$ plot to the one for the state 1000. The probability of obtaining each state for the two noisy simulations (parameters in \cref{Tab:ParameterValuesMasterEquation}) are given in \cref{Tab:ProbStatesFourRoundNoisy}. It can be seen that the probability of the state 0101 to correct shift errors is significantly more affected by the noise than the state 1000 or 1111. To understand why this is the case, it is useful to look at the wave function densities for these three states in the $q$ basis when no noise is present (see \cref{fig:qBasisGoodAndBadStates}). Comparing the wave function densities, it can be observed that the state 1000 has three peaks with similar amplitudes, whereas the states 0101 and 1111 have two smaller peaks compared to the peak at the center. Performing a numerical analysis, it is found that when only damping is present (with all other noise terms including Kerr and non-linear dispersive shift set to zero), damping has a negligible effect on the height of the peaks for all considered states. However performing a numerical simulation with only the Kerr and non-linear dispersive shift terms present, these terms significantly reduce the height of the peaks for the state 0101 and 1010. Therefore, these states are left with one large peak in $q$ with all other peaks close to zero which results in a wave function density with very low resolution in $p$ space (which thus affects the ability for these states to correct shift errors in $p$). Interestingly, the state 1111 is more robust to the Kerr and non-linear dispersive shift contributions since there is a much smaller reduction in the amplitudes of the smaller peaks compared to those for the states 0101 and 1010 (see \cref{fig:KerrOnlyStateComparison}). Furthermore, the states 1000 and 0010 have three large peaks in $q$ and thus even with a reduced amplitude, these states have a higher wave function density resolution in $p$ space.

\section{Conclusion}
\label{sec:conclusion}

In this work we presented a fault-tolerant state preparation protocol for preparing GKP states using phase estimation. In \cref{sec:GoodnessGKP} we provided metrics for comparing how good approximate GKP states are at correcting shift errors in both $q$ and $p$ space. In \cref{sec:FaultToltDefs}, we provided a definition for the $(m,\tilde{\delta})$-Fault-tolerant state preparation of an approximate GKP state where $m$ is the maximum number of allowable faults which can occur on the qubit Hilbert space and $\tilde{\delta}$ is the maximum allowed shift error that can affect the oscillator. Using a non-adaptive phase estimation protocol with one ancilla qubit and one flag qubit, in \cref{Sec:FaultTolerantPhaseEstimation} we showed how the protocol can be made into a $(1,0.06)$-fault-tolerant state preparation of an approximate GKP state. The flag qubit is used to detect damping errors. In addition, it was shown how the adaptive phase estimation protocol of \cref{subsec:FaulFreePhaseEstimation} can not be made fault-tolerant in the presence of measurement readout errors. For the four round non-adaptive phase estimation protocol, 5 of the 16 output states can be used in the presence of measurement readout errors. The total probability of obtaining the accepted states is approximately 0.48. In \cref{sec:CircuitQED}, we considered how the protocol can be implemented in circuit QED. We first provided (to leading order) analytic expressions of the non-linear dispersive shift and Kerr terms during the evolution of the qubit and cavity in the fault-tolerant phase estimation protocol. We used these expressions to find a Gaussian pulse shape that allows one to implement the desired gates. However to mitigate effects due to the non-linear dispersive shift and Kerr terms would require the protocol to be implemented on time scales four orders of magnitude larger than those that were considered in this work. Due to the noise processes afflicting the system, such long time scales would render the protocol impractical. Performing two different simulations for both current and futuristic parameter values found in 2D and 3D cavities, we numerically solved a master equation to study the affects of qubit damping, dephasing and photon loss on the cavity in addition to gate and measurement errors during the protocol. It is shown numerically that 3 of the 5 accepted states (for the four round non-adaptive phase estimation protocol) are much more robust to noise arising from the non-linear dispersive shift and Kerr terms and maintain good error correction capabilities even in the presence of noise. 

The pulse shape used in this work to implement the controlled-displacement gate of the protocol was obtained from the analytical expressions describing the time evolution of the qubit-cavity system. An important direction for future work is to find a pulse shape using methods such as optimal control \cite{OptimalControl1,OptimalControl2} in order to obtain a pulse which can further mitigate the effects from the non-linear dispersive shift and Kerr terms. Since the non-linear dispersive shift and Kerr terms are the dominant source of noise that reduce the error correcting capabilities of approximate GKP states, using optimal control could potentially allow the protocol to be implemented using a larger number of rounds in order to obtain better approximate GKP states. 

The fault-tolerant state preparation of approximate GKP states presented in this work is tailored to protocols that use phase estimation. An interesting direction for future work would be to find fault-tolerant implementations for preparing approximate GKP states that apply to broader schemes such as those found in \cite{GKPStatePrepHome16,GKPStatePrepHomePhotonCatalysis}. In addition, fault-tolerant state preparation protocols for hexagonal GKP codes could be analyzed since these offer better error correction capabilities than GKP codes on a square lattice. It would also be interesting to extend the ideas presented in this work beyond state preparation. For instance, fault-tolerant protocols for the implementation of logical gates would be of great interest.

\section{Acknowledgements}
We thank Barbara Terhal, Daniel Weigand and Daniel Gottesman for useful feedback and discussions. We give special thanks to Zlatko Minev for the many useful discussions regarding circuit QED. Y.S. performed the majority of the numerical simulations. C.C. performed the majority of the analytical calculations, developed the fault-tolerant schemes and wrote the majority of the manuscript. Y.S. and A.C. acknowledges scientific discussions with Fred Chong that took place during the completion of this project. Y.S. is funded in part by EPiQC, an NSF Expedition in
Computing, under grant CCF-1730449 and NSF award STAQ, PFCQC-1818914. Y.S. is also funded in part by the NSF QISE-NET fellowship under grant number 1747426. This work was completed in part with resources provided
by the University of Chicago Research Computing Center.

\newpage 

\bibliographystyle{ieeetr}
\bibliography{bibtex_chamberland}

\begin{thebibliography}{10}

\bibitem{AlbertBosonicCodePerformance18}
V.~V. Albert, K.~Noh, K.~Duivenvoorden, D.~J. Young, R.~T. Brierley,
  P.~Reinhold, C.~Vuillot, L.~Li, C.~Shen, S.~M. Girvin, B.~M. Terhal, and
  L.~Jiang, ``Performance and structure of single-mode bosonic codes,'' {\em
  Phys. Rev. A}, vol.~97, p.~032346, Mar 2018.

\bibitem{GaussianChannel18}
K.~{Noh}, V.~V. {Albert}, and L.~{Jiang}, ``Quantum capacity bounds of gaussian
  thermal loss channels and achievable rates with gottesman-kitaev-preskill
  codes,'' {\em IEEE Transactions on Information Theory}, vol.~65,
  pp.~2563--2582, April 2019.

\bibitem{BarbaraDanielGKP}
B.~M. Terhal and D.~Weigand, ``Encoding a qubit into a cavity mode in circuit
  qed using phase estimation,'' {\em Phys. Rev. A}, vol.~93, p.~012315, Jan
  2016.

\bibitem{ToricGKPBarbara}
C.~Vuillot, H.~Asasi, Y.~Wang, L.~P. Pryadko, and B.~M. Terhal, ``Quantum error
  correction with the toric gottesman-kitaev-preskill code,'' {\em Phys. Rev.
  A}, vol.~99, p.~032344, Mar 2019.

\bibitem{GKPFaultToleranceJapan}
K.~Fukui, A.~Tomita, A.~Okamoto, and K.~Fujii, ``High-threshold fault-tolerant
  quantum computation with analog quantum error correction,'' {\em Phys. Rev.
  X}, vol.~8, p.~021054, May 2018.

\bibitem{NohChambsGKPSurface}
K.~{Noh} and C.~{Chamberland}, ``{Fault-tolerant bosonic quantum error
  correction with the surface-GKP code},'' {\em arXiv e-prints},
  p.~arXiv:1908.03579, Aug 2019.

\bibitem{MagicStateGKP19}
B.~Q. {Baragiola}, G.~{Pantaleoni}, R.~N. {Alexand er}, A.~{Karanjai}, and
  N.~C. {Menicucci}, ``{All-Gaussian universality and fault tolerance with the
  Gottesman-Kitaev-Preskill code},'' {\em arXiv e-prints}, p.~arXiv:1903.00012,
  Feb 2019.

\bibitem{FirstGKPPrep02}
B.~C. Travaglione and G.~J. Milburn, ``Preparing encoded states in an
  oscillator,'' {\em Phys. Rev. A}, vol.~66, p.~052322, Nov 2002.

\bibitem{GlancyKnillCondition}
S.~Glancy and E.~Knill, ``Error analysis for encoding a qubit in an
  oscillator,'' {\em Phys. Rev. A}, vol.~73, p.~012325, Jan 2006.

\bibitem{GKPStatePrepHome16}
J.~Alonso, F.~M. Leupold, Z.~U. Soler, M.~Fadel, M.~Marinelli, B.~C. Keitch,
  V.~Negnevitsky, and J.~P. Home, ``Generation of large coherent states by
  bang-bang control of a trapped-ion oscillator,'' {\em Nature communications},
  vol.~7, no.~11243, 2016.

\bibitem{HomeNatureGKP}
C.~Fluhmann, T.~L. Nguyen, M.~Marinelli, V.~Negnevitsky, K.~Mehta, and J.~P.
  Home, ``Encoding a qubit in a trapped-ion mechanical oscillator,'' {\em
  Nature}, vol.~566, no.~7745, pp.~513--517, 2019.

\bibitem{TaiwanGKPPrep18}
W.-C. {Su}, C.-Y. {Lin}, and S.-T. {Wu}, ``{Encoding qubits into
  harmonic-oscillator modes via quantum walks in phase space},'' {\em arXiv
  e-prints}, p.~arXiv:1808.08722, Aug 2018.

\bibitem{GKPStatePrepHomePhotonCatalysis}
M.~Eaton, R.~Nehra, and O.~Pfister, ``Gottesman-kitaev-preskill state
  preparation by photon catalysis,'' {\em arXiv:quant-ph/1903.01925}, 2019.

\bibitem{ProbFaultGKP19}
T.~Douce, D.~Markham, E.~Kashefi, P.~van Loock, and G.~Ferrini, ``Probabilistic
  fault-tolerant universal quantum computation and sampling problems in
  continuous variables,'' {\em Phys. Rev. A}, vol.~99, p.~012344, Jan 2019.

\bibitem{ShrutiStabilizedCat19}
S.~{Puri}, A.~{Grimm}, P.~{Campagne-Ibarcq}, A.~{Eickbusch}, K.~{Noh},
  G.~{Roberts}, L.~{Jiang}, M.~{Mirrahimi}, M.~H. {Devoret}, and S.~M.
  {Girvin}, ``{Stabilized Cat in Driven Nonlinear Cavity: A Fault-Tolerant
  Error Syndrome Detector},'' {\em arXiv e-prints}, p.~arXiv:1807.09334, Jul
  2018.

\bibitem{AGP06}
P.~Aliferis, D.~Gottesman, and J.~Preskill, ``Quantum accuracy threshold for
  concatenated distance-3 codes,'' {\em Quantum Info. Comput.}, vol.~6,
  pp.~97--165, Mar. 2006.

\bibitem{GKP2001}
D.~Gottesman, A.~Kitaev, and J.~Preskill, ``Encoding a qubit in an
  oscillator,'' {\em Phys. Rev. A}, vol.~64, p.~012310, Jun 2001.

\bibitem{CDT09}
A.~W. Cross, D.~P. Divincenzo, and B.~M. Terhal, ``A comparative code study for
  quantum fault tolerance,'' {\em Quantum Info. Comput.}, vol.~9, pp.~541--572,
  July 2009.

\bibitem{AliferisCross07}
P.~Aliferis and A.~W. Cross, ``Subsystem fault tolerance with the bacon-shor
  code,'' {\em Phys. Rev. Lett.}, vol.~98, p.~220502, May 2007.

\bibitem{CJL16}
C.~Chamberland, T.~Jochym-O'Connor, and R.~Laflamme, ``Thresholds for universal
  concatenated quantum codes,'' {\em Phys. Rev. Lett.}, vol.~117, p.~010501,
  2016.

\bibitem{CJL16b}
C.~Chamberland, T.~Jochym-O'Connor, and R.~Laflamme, ``Overhead analysis of
  universal concatenated quantum codes,'' {\em Phys. Rev. A}, vol.~95,
  p.~022313, Feb 2017.

\bibitem{ChamberlandNeuralNet}
C.~Chamberland and P.~Ronagh, ``Deep neural decoders for near term
  fault-tolerant experiments,'' {\em Quantum Science and Technology}, vol.~3,
  p.~044002, jul 2018.

\bibitem{DispSensorBarbara}
K.~Duivenvoorden, B.~M. Terhal, and D.~Weigand, ``Single-mode displacement
  sensor,'' {\em Phys. Rev. A}, vol.~95, p.~012305, Jan 2017.

\bibitem{Rosenblum266}
S.~Rosenblum, P.~Reinhold, M.~Mirrahimi, L.~Jiang, L.~Frunzio, and R.~J.
  Schoelkopf, ``Fault-tolerant detection of a quantum error,'' {\em Science},
  vol.~361, no.~6399, pp.~266--270, 2018.

\bibitem{CR17v1}
R.~Chao and B.~W. Reichardt, ``Quantum error correction with only two extra
  qubits,'' {\em Phys. Rev. Lett.}, vol.~121, p.~050502, Aug 2018.

\bibitem{CR17v2}
R.~Chao and B.~W. Reichardt, ``Fault-tolerant quantum computation with few
  qubits,'' {\em npj Quantum Information}, vol.~4, 2018.

\bibitem{CB17}
C.~Chamberland and M.~E. Beverland, ``Flag fault-tolerant error correction with
  arbitrary distance codes,'' {\em {Quantum}}, vol.~2, p.~53, Feb. 2018.

\bibitem{TCD18Flag}
T.~Tansuwannont, C.~Chamberland, and D.~Leung, ``Flag fault-tolerant error
  correction for cyclic css codes,'' {\em arXiv:1803.09758}, 2018.

\bibitem{ReichardtFlag18}
B.~W. Reichardt, ``Fault-tolerant quantum error correction for steane's
  seven-qubit color code with few or no extra qubits,'' {\em
  arXiv:quant-ph/1804.06995}, 2018.

\bibitem{CC18MagicFlag}
C.~Chamberland and A.~Cross, ``Fault-tolerant magic state preparation with flag
  qubits,'' {\em arXiv:quant-ph/1811.00566}, 2018.

\bibitem{BlaisFirstPaper}
J.~Gambetta, A.~Blais, M.~Boissonneault, A.~A. Houck, D.~I. Schuster, and S.~M.
  Girvin, ``Quantum trajectory approach to circuit qed: Quantum jumps and the
  zeno effect,'' {\em Phys. Rev. A}, vol.~77, p.~012112, Jan 2008.

\bibitem{BlaisSecondPaper}
M.~Boissonneault, J.~M. Gambetta, and A.~Blais, ``Dispersive regime of circuit
  qed: Photon-dependent qubit dephasing and relaxation rates,'' {\em Phys. Rev.
  A}, vol.~79, p.~013819, Jan 2009.

\bibitem{OptimalControl2}
N.~Ofek, A.~Petrenko, R.~Heeres, P.~Reinhold, Z.~Leghtas, B.~Vlastakis, Y.~Liu,
  L.~Frunzio, S.~M. Girvin, L.~Jiang, M.~Mirrahimi, M.~H. Devoret, and R.~J.
  Schoelkopf, ``Demonstrating quantum error correction that extends the
  lifetime of quantum information,'' {\em arXiv:quant-ph/1602.04768}, 2016.

\bibitem{JOHANSSON20131234}
J.~Johansson, P.~Nation, and F.~Nori, ``Qutip 2: A python framework for the
  dynamics of open quantum systems,'' {\em Computer Physics Communications},
  vol.~184, no.~4, pp.~1234 -- 1240, 2013.

\bibitem{GKPparams1}
K.~S. Chou, J.~Z. Blumoff, C.~S. Wang, P.~C. Reinhold, C.~J. Axline, Y.~Y. Gao,
  L.~Frunzio, M.~H. Devoret, L.~Jiang, and R.~J. Schoelkopf, ``Deterministic
  teleportation of a quantum gate between two logical qubits,'' {\em Nature},
  vol.~561, no.~7723, pp.~1476--4687, 2018.

\bibitem{GKPparams2}
Y.~Y. Gao, B.~J. Lester, K.~S. Chou, L.~Frunzio, M.~H. Devoret, L.~Jiang, S.~M.
  Girvin, and R.~J. Schoelkopf, ``Entanglement of bosonic modes through an
  engineered exchange interaction,'' {\em Nature}, vol.~566, no.~7745,
  pp.~1476--4687, 2019.

\bibitem{LongCavityLifetimes}
M.~Reagor, H.~Paik, G.~Catelani, L.~Sun, C.~Axline, E.~Holland, I.~M. Pop,
  N.~A. Masluk, T.~Brecht, L.~Frunzio, M.~H. Devoret, L.~Glazman, and R.~J.
  Schoelkopf, ``Reaching 10 ms single photon lifetimes for superconducting
  aluminum cavities,'' {\em Applied Physics Letters}, vol.~102, no.~19,
  p.~192604, 2013.

\bibitem{2DCavityLifetimes}
Z.~K. Minev, K.~Serniak, I.~M. Pop, Z.~Leghtas, K.~Sliwa, M.~Hatridge,
  L.~Frunzio, R.~J. Schoelkopf, and M.~H. Devoret, ``Planar multilayer circuit
  quantum electrodynamics,'' {\em Phys. Rev. Applied}, vol.~5, p.~044021, Apr
  2016.

\bibitem{2DCavityLifetimesv2}
T.~Brecht, W.~Pfaff, C.~Wang, Y.~Chu, L.~Frunzio, M.~H. Devoret, and R.~J.
  Schoelkopf, ``Multilayer microwave integrated quantum circuits for scalable
  quantum computing,'' {\em npj Quantum Information}, vol.~2, no.~16002, 2016.

\bibitem{OptimalControl1}
N.~Khaneja, T.~Reiss, C.~Kehlet, T.~Schulte-Herbruggen, and S.~J. Glaser,
  ``Optimal control of coupled spin dynamics: design of nmr pulse sequences by
  gradient ascent algorithms,'' {\em Journal of Magnetic Resonance}, vol.~172,
  pp.~296--305, 2005.

\end{thebibliography}

\clearpage

\appendix 

\section{Shift error arising from amplitude damping before the controlled-displacement gate}
\label{sec:ControlDispAmpDamShiftErr}

\begin{figure}
    \centering
    \includegraphics[height=4.5cm]{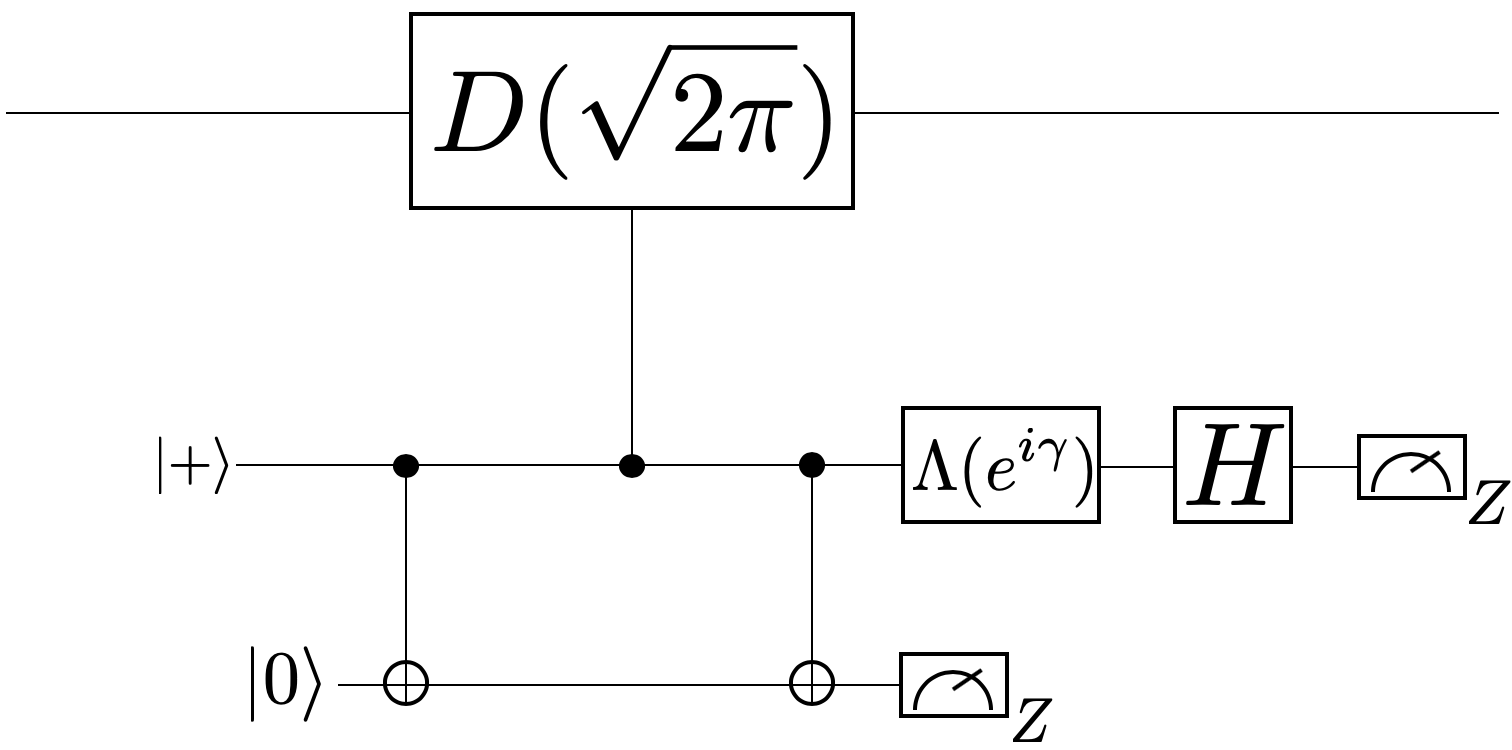}   
    \caption{Controlled-displacement circuit with the flag qubit. The protocol is aborted if the flag qubit measurement is non-trivial.}
    \label{fig:ControlDispAmpDampFig}
\end{figure}

Consider amplitude damping on the $\ket{+}$ ancilla state before the controlled displacement gate as shown in  \cref{fig:ControlDispAmpDampFig}. Let $\ket{\varphi}$ be the input state part of the cavity Hilbert space. Defining $p$ as the damping rate, after the controlled-displacement gate, the state of the system can be written as
\begin{align}
\ket{\psi^{(1)}} = \frac{1}{\sqrt{2}} \Big \{ \ket{00}\ket{\varphi} + \sqrt{1-p}\ket{11} D(\sqrt{2 \pi}) \ket{\varphi} + \sqrt{p} \ket{01} \ket{\varphi}     \Big \}.
\end{align}
Here we have omitted writing the state of the environment interacting with the first ancilla qubit. However, this does not affect the result of the calculations that follow. 

After the second CNOT gate, the state becomes
\begin{align}
\ket{\psi^{(2)}} = \frac{1}{\sqrt{2}} \Big \{ \ket{00}\ket{\varphi} + \sqrt{1-p}\ket{10} D(\sqrt{2 \pi}) \ket{\varphi} + \sqrt{p} \ket{01} \ket{\varphi}     \Big \}.
\end{align}

If the flag qubit is measured as 1, the protocol is aborted. So assume that the flag is measured as 0. In this case the state becomes (tracing out the flag qubit)
\begin{align}
\ket{\psi^{(3)}} = \frac{1}{\sqrt{2 - p}} \Big \{ \ket{0}\ket{\varphi} + \sqrt{1-p}\ket{1} D(\sqrt{2 \pi}) \ket{\varphi}  \Big \}.
\end{align}

Applying the remaining gates in \cref{fig:ControlDispAmpDampFig}, the final state prior to the measurement of the ancilla qubit is
\begin{align}
\ket{\psi^{(4)}} = \frac{1}{\sqrt{2(2 - p)}} \Big \{ \ket{0}(1 + e^{i \gamma}\sqrt{1-p} D(\sqrt{2 \pi}))\ket{\varphi} + \ket{1}(1 - e^{i \gamma}\sqrt{1-p} D(\sqrt{2 \pi}))\ket{\varphi}  \Big \}.
\end{align}

If the measurement of the ancilla is 0, the output state will be
\begin{align}
\ket{\psi^{\text{out,0}}} =  \frac{1}{\sqrt{2 - p}} \Big \{ (1 + e^{i \gamma} \sqrt{1-p} D(\sqrt{2 \pi}) ) \ket{\varphi} \Big \}.
\end{align}
If the measurement of the ancilla is 1, the output state will be
\begin{align}
\ket{\psi^{\text{out,1}}} =  \frac{1}{\sqrt{2 - p}} \Big \{ (1 - e^{i \gamma} \sqrt{1-p} D(\sqrt{2 \pi}) ) \ket{\varphi} \Big \}.
\end{align}

Both outcomes occur with probability $1/2$.

\section{Analytic derivation of the unitary operator describing the evolution of the qubit-cavity system during the implementation of the controlled-$D(\sqrt{2 \pi})$ gate.}
\label{App:NoiseAnalysisSection}

\subsection{Implementation of the controlled-$D(\sqrt{2 \pi})$ gate in the lab frame.}
\label{sec:ControlDispShift}

In this section, we will derive the unitary operator describing the evolution of the qubit-cavity state during the application of the control-displacement gate. In the dispersive regime, the qubit-cavity interaction can be described as:

\begin{align}
H(t) = H_{s} + H_{d}(t),
\label{eq:FullHControllDisp}
\end{align}
where

\begin{align}
H_{s} = \tilde{\omega}_{r} a^{\dagger}a + \tilde{\omega}_{a}Z + \chi a^{\dagger}aZ - \phi (a^{\dagger}a)^{2}Z.
\label{eq:Hst}
\end{align}
and
\begin{align}
H_{d}(t) = \mathcal{E}(t)(a + a^{\dagger}).
\label{eq:Hdt}
\end{align}

In \cref{eq:Hst}, $\phi = \frac{g^4}{\Delta^{3}}$, $\tilde{\omega}_{r} = \omega_{r} + \phi$, $\chi = \frac{g^2}{\Delta} - \phi$ and $\tilde{\omega}_{a} = \frac{\omega_{a} + \chi}{2}$. The term $\phi (a^{\dagger}a)^{2}Z$ corresponds to the non-linear dispersive shift. Note that we have not included the Kerr term $-\frac{K}{2}(a^{\dagger}a)^{2}$. At the end of this section we will modify our results to take into account its effect. We also note that in writing the drive Hamiltonian in \cref{eq:Hdt}, we neglected a term of the form $\lambda X$ (where $\lambda = (g/ \Delta$) and all higher powers in $\lambda$. For parameter values considered in this paper, we found the effect of this term to be negligible. Additionally, we chose a pulse shape which is real valued. 

In \cref{eq:Hdt}, we represent the drive pulse $\mathcal{E}(t)$ as
\begin{align}
\mathcal{E}(t) = \Omega_{x}(t)\cos{(\omega_{d}t)} + \Omega_{y}(t)\sin{(\omega_{d}t)}, 
\end{align}
where $\omega_{d} = \tilde{\omega}_{r} - \chi$ is the drive frequency. 

Since the Hamiltonian does not commute at different times, the unitary operator describing the time evolution under the Hamiltonian of \cref{eq:FullHControllDisp} is given by

\begin{align}
V(0,T) = \mathcal{T} e^{-i \int_{0}^{T}H(t')dt'}.
\label{eq:Unitary1}
\end{align}

The right-hand side of \cref{eq:Unitary1} can be computed using the Suzuki-Trotter decomposition, so that

\begin{align}
V(0,T) &= \Big( \lim_{n \to \infty} \prod_{j = 1}^{n} e^{\frac{-iT}{n}(\tilde{\omega}_{r} + (\chi-\phi a^{\dagger}a) Z)a^{\dagger}a} e^{\frac{-iT}{n} \mathcal{E}(t_{j})(a+a^{\dagger})}   \Big ) e^{-iT \tilde{\omega}_{a} Z} \nonumber \\ 
&=  \Big( \lim_{n \to \infty} \prod_{j = 1}^{n} e^{-i \tilde{t}(\omega_{+} -\phi a^{\dagger}a)a^{\dagger}a} e^{-i \tilde{t}\mathcal{E}(t_{j})(a+a^{\dagger})}   \Big ) e^{-i \tilde{\omega}_{a}T} \ket{0} \bra{0} + \Big( \lim_{n \to \infty} \prod_{j = 1}^{n} e^{-i\tilde{t}(\omega_{-} +\phi a^{\dagger}a)a^{\dagger}a} e^{-i\tilde{t} \mathcal{E}(t_{j})(a+a^{\dagger})}   \Big ) e^{i  \tilde{\omega}_{a}T}\ket{1} \bra{1},
\label{eq:Unitary2}
\end{align}
where $\tilde{t} \equiv T/n$, $\omega_{\pm} \equiv \tilde{\omega_{r}} \pm \chi$.

Now, the two terms on the right-hand side of \cref{eq:Unitary2} can be decomposed as 

\begin{align}
&\lim_{n \to \infty} \prod_{j = 1}^{n} e^{-i \tilde{t}(\omega_{+} -\phi a^{\dagger}a)a^{\dagger}a} e^{-i \tilde{t}\mathcal{E}(t_{j})(a+a^{\dagger})} \nonumber \\
&= \lim_{n \to \infty} R_{n}^{n} (R_{n}^{-n} D_{t_n} R_{n}^{n}) \cdots (R_{n}^{-2} D_{t_2} R_{n}^{2})(R_{n}^{-1} D_{t_1} R_{n}^{1}),
\label{eqRnproducts}
\end{align}
where
\begin{align}
R_{n} = e^{-i \tilde{t}(\omega_{+} -\phi a^{\dagger}a)a^{\dagger}a},
\label{eq:RnDef}
\end{align}
and
\begin{align}
D_{t_j} = e^{-i \tilde{t}\mathcal{E}(t_{j})(a+a^{\dagger})}.
\label{eq:DtjDef}
\end{align}

Hence from the above, we see that we need to compute terms of the form
\begin{align}
R_{n}^{-k} D_{t_k} R_{n}^{k} = e^{i \tilde{t}k(\omega_{+} -\phi a^{\dagger}a)a^{\dagger}a} e^{-i \tilde{t}\mathcal{E}(t_{k})(a+a^{\dagger})} e^{-i \tilde{t}k(\omega_{+} -\phi a^{\dagger}a)a^{\dagger}a}.
\label{eq:ProductToCompute}
\end{align}

In order to compute the products appearing in \cref{eq:ProductToCompute}, we will use the identity $Ae^{B}A^{-1} = e^{ABA^{-1}}$. Defining,
\begin{align}
H' = k(\omega_{+} - \phi a^{\dagger} a) a^{\dagger} a,
\label{eq:HPrimeDef}
\end{align}
and
\begin{align}
A_{\tilde{t}} = e^{-i \tilde{t} \mathcal{E}(t_k)(a + a^{\dagger})},
\label{eq:AtildeDef}
\end{align}
with
\begin{align}
A(\tilde{t}) = e^{iH' \tilde{t}} A_{\tilde{t}} e^{-iH' \tilde{t}}
\label{eq:AdefSandwhich}
\end{align}
we have that
\begin{align}
A(\tilde{t}) = \text{exp}\Big (-i\tilde{t} \mathcal{E}(t_{k}) e^{iH' \tilde{t}}(a + a^{\dagger})e^{-iH' \tilde{t}}   \Big ),
\label{eq:HeisenberAtildeT}
\end{align}
The term $e^{iH' \tilde{t}}(a + a^{\dagger})e^{-iH' \tilde{t}}$ in \cref{eq:HeisenberAtildeT} can be computed using Heisenberg's equation of motion. We obtain

\begin{align}
\frac{d a(\tilde{t})}{d \tilde{t}} = i[H',a(\tilde{t})],
\end{align}
with
\begin{align}
[H',a] = k(\phi ' + 2\phi a^{\dagger}a)a,
\end{align}
where we defined
\begin{align}
\phi ' \equiv \phi - \omega_{+}.
\end{align}

We thus have the following set of coupled differential equations:
\begin{align}
\frac{d a(\tilde{t})}{d \tilde{t}} = ik(\phi ' +2\phi a^{\dagger}(\tilde{t})a(\tilde{t}))a(\tilde{t}),
\label{eq:CoupledDiff1}
\end{align}
and
\begin{align}
\frac{d a^{\dagger}(\tilde{t})}{d \tilde{t}} = -ik a^{\dagger}(\tilde{t})(\phi ' +2\phi a^{\dagger}(\tilde{t})a(\tilde{t})),
\label{eq:CoupledDiff2}
\end{align}
The solutions to \cref{eq:CoupledDiff1,eq:CoupledDiff2} are given by
\begin{align}
a(\tilde{t}) = a e^{ik\tilde{t} (\phi ' +2\phi(a^{\dagger}a - 1))},
\label{eq:SolDiff1}
\end{align}
and
\begin{align}
a^{\dagger}(\tilde{t}) = e^{-ik\tilde{t} (\phi ' +2\phi(a^{\dagger}a - 1))}a^{\dagger}.
\label{eq:SolDiff2}
\end{align}

To see this, we take a derivative of \cref{eq:SolDiff1} to obtain
\begin{align}
\frac{d a(\tilde{t})}{d \tilde{t}} = ik \phi ' a(\tilde{t}) + 2ik \phi a(\tilde{t})(a^{\dagger}a - 1).
\label{eq:DerivativeOfSol1}
\end{align}
Comparing \cref{eq:CoupledDiff1,eq:DerivativeOfSol1}, we must have that 
\begin{align}
a(\tilde{t})(a^{\dagger}a - 1) = a^{\dagger}(\tilde{t})a^{2}(\tilde{t}).
\end{align}
Using \cref{eq:SolDiff1,eq:SolDiff2} and defining $H_{te} \equiv -kt(\phi ' + 2\phi(a^{\dagger}a - 1))$, we have that
\begin{align}
a^{\dagger}(\tilde{t})a^{2}(\tilde{t}) &= e^{i H_{te} \tilde{t}} a^{\dagger}e^{-i H_{te} \tilde{t}} e^{i H_{te} \tilde{t}}a e^{-i H_{te} \tilde{t}} a e^{-i H_{te} \tilde{t}} \nonumber \\
&= a^{\dagger}e^{-2i\phi k \tilde{t}} a e^{2i\phi k \tilde{t}} a e^{-i H_{te} \tilde{t}} \nonumber \\
&= a^{\dagger}a ae^{-i H_{te} \tilde{t}} \nonumber \\
&= a^{\dagger} a a(\tilde{t}),
\end{align}
as desired. A similar calculation can be done for $a^{\dagger}(\tilde{t})$. Hence we have that
\begin{align}
A(\tilde{t}) = \text{exp} \Big(-i \tilde{t} \mathcal{E}(t_{k})(ae^{i\Phi_{k}\tilde{t}} + e^{-i\Phi_{k}\tilde{t}}a^{\dagger})      \Big),
\label{eq:Delatildet}
\end{align}
for
\begin{align}
\Phi_{k} = k(\phi' + 2\phi(a^{\dagger}a - 1)).
\end{align}

Writing $V(0,T)$ (\cref{eq:Unitary2}) as 
\begin{align}
V(0,T) = V_{+}(0,T) \ket{0}\bra{0} + V_{-}(0,T)\ket{1}\bra{1},
\end{align}
using \cref{eq:Delatildet} and the fact that $R_{n}^{n} = e^{-iT(\omega_{+}-\phi a^{\dagger}a)a^{\dagger}a} \equiv R_{+}(T)$, we can write 
\begin{align}
V_{+}(0,T) &= R_{+}(T) \lim_{n \to \infty} \prod_{k=1}^{n} \text{exp} \Big \{A_{k}a^{\dagger} -aA_{k}^{\dagger} \Big \},
\label{eqVPlusGreatlySimplified}
\end{align}
where we defined
\begin{align}
A_{k} \equiv -\frac{iT}{n} \mathcal{E}(t_k)e^{\frac{-iT}{n}\Phi_{k}}.
\label{eq:AkDefOperator}
\end{align}

Notice that we can write $V_{+}(0,T)$ as products of displacements $D(A_{k})$. However, now $A_{k}$ is an operator instead of a complex number. In order to compute the products in \cref{eqVPlusGreatlySimplified}, we will need to obtain a relation for terms of the form $D(A_k)D(A_j)$ (where $A_k$ is given in \cref{eq:AkDefOperator}). Since $A_k$ is proportional to $T/n$ and remembering that we will take the limit where $n \to \infty$, using Baker-Campbell-Hausdorff, we have the exact relation
\begin{align}
\exp{ \Big \{ A_ka^{\dagger} - aA_{k}^{\dagger} + A_ja^{\dagger} - aA_j \Big \}} = D(A_k)D(A_j)\exp{\Big \{ -\frac{1}{2}[A_ka^{\dagger} - aA_{k}^{\dagger},A_ja^{\dagger} - aA_j] \Big \} }
\label{eq:DisplaceBaker}
\end{align}
The commutator on the right hand side of \cref{eq:DisplaceBaker} has four terms
\begin{align}
[A_ka^{\dagger} - aA_{k}^{\dagger},A_ja^{\dagger} - aA_j] = [A_k a^{\dagger},A_j a^{\dagger}] - [A_ka^{\dagger},aA_j^{\dagger}]
-[aA_k^{\dagger},A_ja^{\dagger}] + [aA_k^{\dagger},aA_j^{\dagger}].
\end{align}
As we will show, two of these terms vanish when taking the limit $n \to \infty$.

First, we compute
\begin{align}
[A_k a^{\dagger},A_j a^{\dagger}] = \Big( \frac{-iT}{n} \Big)^2 \mathcal{E}(t_k)\mathcal{E}(t_j)[e^{-\frac{iT}{n}\Phi_k}a^{\dagger},e^{-\frac{iT}{n}\Phi_j}a^{\dagger}],
\label{eq:BigCommAdagAdag1}
\end{align}
with 
\begin{align}
[e^{-\frac{iT}{n}\Phi_k}a^{\dagger},e^{-\frac{iT}{n}\Phi_j}a^{\dagger}] &= e^{-\frac{iT}{n}\Phi_k}a^{\dagger}e^{-\frac{iT}{n}\Phi_j}a^{\dagger} - e^{-\frac{iT}{n}\Phi_j}a^{\dagger}e^{-\frac{iT}{n}\Phi_k}a^{\dagger} \nonumber \\
& = e^{-\frac{iT}{n}\Phi_k}a^{\dagger}e^{\frac{iT}{n}\Phi_k}e^{-\frac{iT}{n}\Phi_{(k+j)}}a^{\dagger}e^{\frac{iT}{n}\Phi_{(k+j)}}e^{-\frac{iT}{n}\Phi_{k+j}} - e^{-\frac{iT}{n}\Phi_j}a^{\dagger}e^{\frac{iT}{n}\Phi_j}e^{-\frac{iT}{n}\Phi_{(k+j)}}a^{\dagger}e^{\frac{iT}{n}\Phi_{(k+j)}}e^{-\frac{iT}{n}\Phi_{(k+j)}}.
\label{eq:OneOfMany}
\end{align}
Now, terms such as $e^{-\frac{iT}{n}\Phi_j}a^{\dagger}e^{\frac{iT}{n}\Phi_j}$ can be computed by defining 
\begin{align}
H_{\text{temp}} \equiv -j\phi ' (1+\tilde{\delta}(a^{\dagger}a-1)),
\end{align}
and invoking Heisenberg's equation of motion. Using $[H_{\text{temp}},a^{\dagger}] = -j\phi ' \tilde{\delta}a^{\dagger}$ and solving the differential equation, we obtain
\begin{align}
e^{-\frac{iT}{n}\Phi_j}a^{\dagger}e^{\frac{iT}{n}\Phi_j} = e^{-\frac{iT}{n} j \phi ' \tilde{\delta}}a^{\dagger}.
\label{eq:CommutatorLove}
\end{align}

Using the result of \cref{eq:CommutatorLove} into \cref{eq:OneOfMany}, we have
\begin{align}
[e^{-\frac{iT}{n}\Phi_k}a^{\dagger},e^{-\frac{iT}{n}\Phi_j}a^{\dagger}]  &= e^{\frac{-iT}{n} k \phi ' \tilde{\delta}}a^{\dagger}e^{\frac{-iT}{n} (k+j) \phi ' \tilde{\delta}}a^{\dagger}e^{\frac{-iT}{n}\Phi_{(k+j)}} - e^{\frac{-iT}{n} j \phi ' \tilde{\delta}}a^{\dagger}e^{\frac{-iT}{n} (j+k) \phi ' \tilde{\delta}}e^{\frac{-iT}{n}\Phi_{(k+j)}} \nonumber \\
& = e^{\frac{-iT}{n}(j+k)\phi ' \tilde{\delta}}(e^{\frac{-iT}{n}k\phi ' \tilde{\delta}} - e^{\frac{-iT}{n}j\phi ' \tilde{\delta}})(a^{\dagger})^2e^{\frac{-iT}{n}\Phi_{(k+j)}}
\label{eq:Int1234}
\end{align}
Using $\phi ' \tilde{\delta} = 2 \phi$ and expanding \cref{eq:Int1234} to leading order in $\phi$, we obtain
\begin{align}
[e^{-\frac{iT}{n}\Phi_k}a^{\dagger},e^{-\frac{iT}{n}\Phi_j}a^{\dagger}]  & \approx (1-\frac{2iT}{n}(j+k)\phi)(\frac{-2iT}{n}\phi)(a^{\dagger})^2e^{\frac{-iT}{n}\Phi_{(k+j)}} \nonumber \\
& \approx \frac{-2iT}{n}(k-j)\phi(a^{\dagger})^2e^{\frac{-iT}{n}\Phi_{(k+j)}}.
\label{eq:CommAdagAdag11}
\end{align}

Inserting \cref{eq:CommAdagAdag11} into \cref{eq:BigCommAdagAdag1}, we obtain
\begin{align}
[A_k a^{\dagger},A_j a^{\dagger}] =2\Big (\frac{-iT}{n}  \Big )^{3} (k-j) \phi \mathcal{E}(t_k)\mathcal{E}(t_j)(a^{\dagger})^2e^{\frac{-iT}{n}\Phi_{(k+j)}} + \mathcal{O}(\phi^2).
\label{eq:AdagAdagFinalResult}
\end{align}
A similar calculation shows that
\begin{align}
[aA_{k}^{\dagger},aA_{j}^{\dagger}] = 2\Big (\frac{iT}{n}  \Big )^{3} (k-j) \phi \mathcal{E}(t_k)\mathcal{E}(t_j)e^{\frac{iT}{n}\Phi_{(k+j)}}a^2 + \mathcal{O}(\phi^2).
\label{eq:AAFinalResult}
\end{align}

Next we compute the cross terms. We first have
\begin{align}
[A_ka^{\dagger},aA_j^{\dagger}] = \Big(\frac{iT}{n} \Big)^2 \mathcal{E}(t_k)\mathcal{E}(t_j)[e^{\frac{-iT}{n} \Phi_{k}}a^{\dagger},ae^{\frac{iT}{n} \Phi_{j}}],
\label{eq:AwesomeAdagDdog}
\end{align}
with
\begin{align}
[e^{\frac{-iT}{n} \Phi_{k}}a^{\dagger},ae^{\frac{iT}{n} \Phi_{j}}] &= e^{\frac{-iT}{n} \Phi_{k}}a^{\dagger}ae^{\frac{iT}{n} \Phi_{j}} - ae^{\frac{iT}{n} \Phi_{j}}e^{\frac{-iT}{n} \Phi_{k}}a^{\dagger} \nonumber \\
&= e^{\frac{-iT}{n} \Phi_{k}}a^{\dagger}e^{\frac{iT}{n} \Phi_{k}}e^{\frac{-iT}{n} \Phi_{k}}ae^{\frac{iT}{n} \Phi_{k}}e^{\frac{-iT}{n} \Phi_{k}}e^{\frac{iT}{n} \Phi_{j}} - ae^{\frac{-iT}{n} \Phi_{(k-j)}}a^{\dagger} \nonumber \\
&= a^{\dagger}ae^{\frac{-iT}{n} \Phi_{(k-j)}} - ae^{\frac{-iT}{n} \Phi_{(k-j)}} a^{\dagger}e^{\frac{iT}{n} \Phi_{(k-j)}}e^{\frac{-iT}{n} \Phi_{(k-j)}} \nonumber \\
&= (a^{\dagger}a - aa^{\dagger}e^{\frac{-2iT}{n} \phi(k-j)})e^{\frac{-iT}{n} \Phi_{(k-j)}} \nonumber \\
&= (a^{\dagger}a(1-e^{\frac{-2iT}{n} \phi(k-j)}) - e^{\frac{-2iT}{n} \phi(k-j)})e^{\frac{-iT}{n} \Phi_{(k-j)}}.
\label{eq:AdagAInti870}
\end{align}

Expanding the term proportional to $a^{\dagger}a$ to leading order in $\phi$, \cref{eq:AdagAInti870} becomes
\begin{align}
[e^{\frac{-iT}{n} \Phi_{k}}a^{\dagger},ae^{\frac{iT}{n} \Phi_{j}}] \approx \Big ( \frac{2iT}{n}\phi a^{\dagger}a(k-j) - e^{\frac{-2iT}{n} \phi(k-j)}    \Big)e^{\frac{-iT}{n} \Phi_{(k-j)}}.
\label{eq:TempAdagA778}
\end{align}

Inserting \cref{eq:TempAdagA778} into \cref{eq:AwesomeAdagDdog}, we obtain 
\begin{align}
[A_ka^{\dagger},aA_j^{\dagger}]  \approx 2\Big (\frac{iT}{n} \Big )^{3} \phi (k-j) \mathcal{E}(t_k)\mathcal{E}(t_j) a^{\dagger}a e^{\frac{-iT}{n}\Phi_{(k-j)}} - \Big(\frac{iT}{n} \Big)^{2}\mathcal{E}(t_k)\mathcal{E}(t_j) e^{\frac{-2iT}{n}\phi (k-j)}e^{\frac{-iT}{n}\Phi_{(k-j)}}.
\label{eq:LastBigLove}
\end{align}

The last commutator to compute is
\begin{align}
[aA_k^{\dagger},A_ja^{\dagger}] = \Big (\frac{-iT}{n} \Big)^{2}\mathcal{E}(t_k)\mathcal{E}(t_j)[ae^{\frac{iT}{n}\Phi_k},e^{\frac{-iT}{n}\Phi_j a^{\dagger}}],
\label{eq:AAdagBeforeFinal}
\end{align}
with 
\begin{align}
[ae^{\frac{iT}{n}\Phi_k},e^{\frac{-iT}{n}\Phi_j}a^{\dagger}] &= a e^{\frac{iT}{n} \Phi_{(k-j)}}a^{\dagger} - e^{\frac{-iT}{n} \Phi_{j}}a^{\dagger}ae^{\frac{iT}{n} \Phi_{k}} \nonumber \\
&= e^{\frac{iT}{n} \Phi_{(k-j)}}e^{\frac{-iT}{n} \Phi_{(k-j)}}ae^{\frac{iT}{n} \Phi_{(k-j)}}a^{\dagger} - e^{\frac{iT}{n} \Phi_{(k-j)}}e^{\frac{-iT}{n} \Phi_{k}}a^{\dagger}e^{\frac{iT}{n} \Phi_{k}}e^{\frac{-iT}{n} \Phi_{k}}ae^{\frac{iT}{n} \Phi_{k}} \nonumber \\
&= e^{\frac{iT}{n} \Phi_{(k-j)}}(a^{\dagger}a(e^{\frac{2iT}{n} \phi (k-j)}-1) + e^{\frac{2iT}{n} \phi (k-j)}) \nonumber \\
& \approx e^{\frac{iT}{n} \Phi_{(k-j)}} \Big (\frac{2iT}{n} \phi (k-j) a^{\dagger}a +  e^{\frac{2iT}{n} \phi (k-j)}  \Big)
\end{align}
so that 
\begin{align}
[aA_k^{\dagger},A_ja^{\dagger}] \approx 2\Big ( \frac{iT}{n} \Big)^{3} \phi (k-j) \mathcal{E}(t_k)\mathcal{E}(t_j) e^{\frac{iT}{n} \Phi_{(k-j)}}a^{\dagger}a + \Big(\frac{-iT}{n} \Big)^{2} \mathcal{E}(t_k) \mathcal{E}(t_j)e^{\frac{iT}{n} \Phi_{(k-j)}} e^{\frac{2iT}{n} \phi(k-j)}
\label{eq:LastBigBoy}
\end{align}

Inserting \cref{eq:AdagAdagFinalResult,eq:AAFinalResult,eq:LastBigLove,eq:LastBigBoy} into \cref{eq:DisplaceBaker}, we have
\begin{align}
D(A_k + A_j) &= D(A_k)D(A_j) \text{exp} \Big [-\frac{1}{2} \Big \{ 2\Big (\frac{-iT}{n}  \Big )^{3} (k-j) \phi \mathcal{E}(t_k)\mathcal{E}(t_j)(a^{\dagger})^2e^{\frac{-iT}{n}\Phi_{(k+j)}} \nonumber \\
& + 2\Big (\frac{iT}{n}  \Big )^{3} (k-j) \phi \mathcal{E}(t_k)\mathcal{E}(t_j)e^{\frac{iT}{n}\Phi_{(k+j)}}a^2 - 2\Big (\frac{iT}{n} \Big )^{3} \phi (k-j) \mathcal{E}(t_k)\mathcal{E}(t_j) a^{\dagger}a e^{\frac{-iT}{n}\Phi_{(k-j)}} \nonumber \\
& +\Big(\frac{iT}{n} \Big)^{2}\mathcal{E}(t_k)\mathcal{E}(t_j) e^{\frac{-2iT}{n}\phi (k-j)}e^{\frac{-iT}{n}\Phi_{(k-j)}} -2\Big ( \frac{iT}{n} \Big)^{3} \phi (k-j) \mathcal{E}(t_k)\mathcal{E}(t_j) e^{\frac{iT}{n} \Phi_{(k-j)}}a^{\dagger}a \nonumber \\
& - \Big(\frac{-iT}{n} \Big)^{2} \mathcal{E}(t_k) \mathcal{E}(t_j)e^{\frac{iT}{n} \Phi_{(k-j)}} e^{\frac{2iT}{n} \phi(k-j)} \Big \} \Big ].
\label{eq:TrueLove}
\end{align}

Since the product $D(A_n) D(A_{n-1}) \cdots D(A_2)D(A_1)$ will produce sums in the exponents proportional to $n^2$, all terms in \cref{eq:TrueLove} proportional to $(\frac{T}{n})^3$ will vanish in the limit where $n \to \infty$. Hence \cref{eq:TrueLove} simplifies to
\begin{align}
D(A_k + A_j) &= D(A_k)D(A_j) \text{exp} \Big [i \mathcal{E}(t_k) \mathcal{E}(t_j) \Big (\frac{T}{n} \Big)^2 \Big \{\frac{e^{\frac{-2iT}{n} \phi(k-j)}e^{\frac{-iT}{n}\Phi_{(k-j)}} - e^{\frac{2iT}{n} \phi(k-j)}e^{\frac{iT}{n}\Phi_{(k-j)}}}{2i}  \Big \}  \Big ] \nonumber \\
& = D(A_k)D(A_j) \text{exp} \Big [-i \mathcal{E}(t_k) \mathcal{E}(t_j) \Big (\frac{T}{n} \Big)^2 \sin{\Big (\frac{T}{n}\overline{\Phi}_{(k-j)}\Big )}  \Big ],
\label{eq:ProdDisplacedSin}
\end{align}
where $\overline{\Phi}_{(k-j)}$ is defined as 
\begin{align}
\overline{\Phi}_{(k-j)} \equiv (k-j)(\phi -\omega_{+} + 2\phi a^{\dagger}a).
\end{align}

We thus have that the product of the modified displacement operators is given by
\begin{align}
D(A_k)D(A_j) = D(A_k + A_j) \text{exp} \Big [i \mathcal{E}(t_k) \mathcal{E}(t_j) \Big (\frac{T}{n} \Big)^2 \sin{\Big ( \frac{T}{n}\overline{\Phi}_{(k-j)}\Big )}  \Big ].
\label{eq:FinalProdDispModifiedLove}
\end{align}

We now wish to compute
\begin{align}
P_n \equiv D(A_n) D(A_{n-1}) D(A_{n-2}) \cdots D(A_2)D(A_1).
\end{align}
First notice that for any operators $A$ and $B$, to leading order $e^{A}e^{B} = e^{B}e^{A}e^{[A,B]}$. Now for terms of the form $\text{exp} \Big [ i\mathcal{E}(t_k)\mathcal{E}(t_j) \Big ( \frac{T}{n} \Big)^2 \sin\Big ( \frac{T}{n} \overline{\Phi}_{(k-j)}   \Big ) \Big ] D(\sum_{m}A_m)$, the commutator of the two exponents will be proportional to $(\frac{T}{n})^3$ which will vanish in the limit where $n \to \infty$. Hence, we can commute all terms $\text{exp} \Big [ i\mathcal{E}(t_k)\mathcal{E}(t_j) \Big ( \frac{T}{n} \Big)^2 \sin\Big ( \frac{T}{n} \overline{\Phi}_{(k-j)}   \Big ) \Big ]$ to the right hand side of $P_n$. Hence we have
\begin{align}
P_n = D \Big( \sum_{k=1}^{n} A_k \Big) \text{exp} \Big [ i \sum_{k < j}^{n}  \frac{T^2}{n^2} \mathcal{E}(t_k)\mathcal{E}(t_j) \sin\Big ( \frac{T}{n} \overline{\Phi}_{(k-j)}   \Big ) \Big ].
\end{align}

Now, taking the limit where $n \to \infty$ of $P_n$, we obtain
\begin{align}
 \lim_{n \to \infty} P_n &= D \Big ( \lim_{n \to \infty} \sum_{k=1}^{n} \Big (\frac{-iT}{n} \Big) \mathcal{E}(t_k) e^{\frac{-iT}{n} \Phi_k}     \Big ) \text{exp} \Big [ i  \lim_{n \to \infty} \sum_{k < j}^{n}  \frac{T^2}{n^2} \mathcal{E}(t_k)\mathcal{E}(t_j) \sin\Big ( \frac{T}{n} \overline{\Phi}_{(k-j)}   \Big ) \Big ] \nonumber \\
 &= D \Big (-i \int_{0}^{T} \mathcal{E}(t) e^{-i \Phi t}dt \Big ) \text{exp} \Big [ -i \int_{0}^{T}dt \int_{t}^{T}dt' \mathcal{E}(t) \mathcal{E}(t') \sin{\Big (\overline{\Phi}(t' -t) \Big )}   \Big ]
\end{align}

We conclude that
\begin{align}
V_{+}(0,T) = R_{+}(T)D(A_{+})e^{iB_{+}},
\label{eq:FinalVPlusbeforExpand}
\end{align}
where
\begin{align}
R_{+}(T) = e^{-iT(\omega_{+}-\phi a^{\dagger}a)a^{\dagger}a},
\label{eq:FinalRRbeforExpand}
\end{align}
\begin{align}
A_{+} = -i \int_{0}^{T} \mathcal{E}(t) e^{-i \Phi t}dt,
\label{eq:FinalAbeforExpand}
\end{align}
and
\begin{align}
B_{+} = - \int_{0}^{T}dt \int_{t}^{T}dt' \mathcal{E}(t) \mathcal{E}(t') \sin{\Big (\overline{\Phi}(t' -t)\Big )}.
\label{eq:FinalBbeforExpand}
\end{align}

Hence to conclude, the unitary evolution of the Hamiltonian described in \cref{eq:FullHControllDisp} is given by
\begin{align}
V(0,T) = R_{+}(T)D(A_{+})e^{iB_{+}}e^{-i \tilde{\omega}_{a}T} \ket{0} \bra{0} + R_{-}(T)D(A_{-})e^{iB_{-}}e^{i \tilde{\omega}_{a}T} \ket{1} \bra{1}.
\end{align}

Recall that in deriving the expression for $V_{+}(0,T)$, in several steps of the calculation we expanded to leading order in $\phi$. Hence the expressions obtained in \cref{eq:FinalRRbeforExpand,eq:FinalAbeforExpand,eq:FinalBbeforExpand} are only valid to leading order in $\phi$. Hence to leading order in $\phi$, we have
\begin{align}
R_{\pm}(T) = e^{-iT \omega_{\pm} a^{\dagger}a}(1 \pm iT \phi (a^{\dagger}a)^2) + \mathcal{O}(\phi^2),
\label{eq:FinalRforReal1}
\end{align}
\begin{align}
A_{\pm} = -i \int_{0}^{T} \mathcal{E}(t) e^{i \omega_{\pm} t}dt \mp \phi(2a^{\dagger}a-1)\int_{0}^{T} \mathcal{E}(t)te^{i \omega_{\pm}t}dt + \mathcal{O}(\phi^2),
\label{eq:FinalAforReal2}
\end{align}
and
\begin{align}
B_{\pm} = \int_{0}^{T}dt \int_{t}^{T}dt' \mathcal{E}(t) \mathcal{E}(t') \sin{\Big (\omega_{\pm}(t' -t)\Big )} \pm \phi(2a^{\dagger}a + 1)\int_{0}^{T}dt \int_{t}^{T}dt' \mathcal{E}(t) \mathcal{E}(t')\cos{\Big (\omega_{\pm}(t' -t)  \Big)}(t'-t) + \mathcal{O}(\phi^2).
\label{eq:FinalBforReal3}
\end{align}

\subsection{Including the Kerr non-linearity}
\label{sec:KerSec}

When including the Kerr non-linearity, the Hamiltonian during the control-displacement gate is now given by
\begin{align}
H(t) = \tilde{\omega}_r a^{\dagger}a + \tilde{\omega}_a Z + \chi a^{\dagger}aZ - \phi(a^{\dagger}a)^2 Z -\frac{K}{2}(a^{\dagger}a)^2.
\end{align}
In this case, we have that
\begin{align}
V(0,T) &=  \Big( \lim_{n \to \infty} \prod_{j = 1}^{n} e^{-i \tilde{t}(\omega_{+} -(\phi + \frac{K}{2})a^{\dagger}a)a^{\dagger}a} e^{-i \tilde{t}\mathcal{E}(t_{j})(a+a^{\dagger})}   \Big ) e^{-i \tilde{\omega}_{a}T} \ket{0} \bra{0} \nonumber \\
& + \Big( \lim_{n \to \infty} \prod_{j = 1}^{n} e^{-i\tilde{t}(\omega_{-} +(\phi - \frac{K}{2}) a^{\dagger}a)a^{\dagger}a} e^{-i\tilde{t} \mathcal{E}(t_{j})(a+a^{\dagger})}   \Big ) e^{i  \tilde{\omega}_{a}T}\ket{1} \bra{1},
\label{eq:Unitary22}
\end{align}

Hence, we see that the analysis of \cref{sec:ControlDispShift} is exactly the same, with $\phi \to \phi \pm \frac{K}{2}$ in $R_{\pm}(T)$, $A_{\pm}$ and $B_{\pm}$. 

\subsection{Controlled-displacement gate in the rotating frame}

In this section, we consider the same Hamiltonian as in \cref{eq:FullHControllDisp} but with  a drive term of the form
\begin{align}
H_{d}(t) = \mathcal{E}(t) a^{\dagger} e^{-i \omega_{d} t} + \mathcal{E}^{*}(t) a e^{i \omega_{d} t}.
\label{eq:RotatingFrameHam11}
\end{align}

We will go into the rotating frame of both the qubit and the cavity. We define 

\begin{align}
H_{s}' = \tilde{\omega}_{r} a^{\dagger}a + \tilde{\omega}_{a}Z.
\end{align}

With $U(t) = e^{iH_{s}' t}$ and applying the transformation $U(t) H(t) U^{-1}(t) - i U(t) \frac{\partial }{\partial t}U^{-1}(t) \equiv H_{R}(t)$ to the Hamiltonian in \cref{eq:FullHControllDisp} with $H_{d}(t)$ given in \cref{eq:RotatingFrameHam11}, we obtain

\begin{align}
H_{R}(t) = \chi Z a^{\dagger}a -\phi (a^{\dagger}a)^{2}Z + \mathcal{E}(t) a^{\dagger} e^{i(\tilde{\omega}_{r} - \omega_{d})t} + \mathcal{E}^{*}(t) a e^{-i(\tilde{\omega}_{r} - \omega_{d})t}
\label{eq:HrotFrequency}
\end{align}
The frequency dependence in the drive term of \cref{eq:HrotFrequency} can be eliminated by choosing $\omega_{d} = \tilde{\omega}_{r}$. Again, we wish to compute the unitary evolution 
\begin{align}
V_{R}(0,T) = \mathcal{T} e^{-i \int_{0}^{T}H_{R}(t')dt'}.
\label{eq:UnitaryRotatingFrame}
\end{align}

Using the Suzuki-Trotter decomposition
\begin{align}
V_{R}(0,T) &=  \Big( \lim_{n \to \infty} \prod_{j = 1}^{n} e^{-i \tilde{t}(\omega_{+} - \phi a^{\dagger}a)a^{\dagger}a} e^{-i \tilde{t}(\mathcal{E}(t_j) a^{\dagger}  + \mathcal{E}^{*}(t_j) a)}   \Big ) \ket{0} \bra{0} \nonumber \\
& + \Big( \lim_{n \to \infty} \prod_{j = 1}^{n} e^{-i \tilde{t}(\omega_{-} + \phi a^{\dagger}a)a^{\dagger}a} e^{-i \tilde{t}(\mathcal{E}(t_j) a^{\dagger} + \mathcal{E}^{*}(t_j) a)}   \Big ) \ket{1} \bra{1},
\label{eq:UnitaryRot2}
\end{align}
where now we have that $\omega_{\pm} = \pm \chi$.

Comparing \cref{eq:UnitaryRot2} to \cref{eq:Unitary2}, we see that we can follow the same steps as in \cref{sec:ControlDispShift} by using the new values for $\omega_{\pm}$ and replacing $\mathcal{E}$ by $\mathcal{E}^{*}$ in the conjugate expressions. Doing so, we obtain 

\begin{align}
V_{R}(0,T) &= R_{+}(T)D(A_{+})e^{iB_{+}} \ket{0}\bra{0} +
R_{-}(T)D(A_{-})e^{iB_{-}} \ket{1}\bra{1},
\label{eq:FinalUnitaryEvol}
\end{align}
where
\begin{align}
R_{\pm}(T) = e^{-iT \omega_{\pm} a^{\dagger}a}(1 \pm iT \phi (a^{\dagger}a)^{2}),
\end{align}

\begin{align}
A_{\pm} = -i \int_{0}^{T} \mathcal{E}(t) e^{i \omega_{\pm} t}dt \mp \phi(2a^{\dagger}a-1)\int_{0}^{T} \mathcal{E}(t)te^{i\omega_{\pm}t}dt + \mathcal{O}(\phi^2),
\end{align}

\begin{align}
B_{\pm} &= -\frac{i}{2}\int_{0}^{T}dt \int_{t}^{T}dt' \Big ( \mathcal{E}^{*}(t') \mathcal{E}(t) e^{i \omega_{\pm} (t' - t)} -   \mathcal{E}(t') \mathcal{E}^{*}(t) e^{-i \omega_{\pm} (t' - t)} \Big ) \nonumber \\  &\pm \frac{\phi(2a^{\dagger}a + 1)}{2}\int_{0}^{T}dt \int_{t}^{T}dt' \Big ( \mathcal{E}^{*}(t') \mathcal{E}(t) e^{i \omega_{\pm} (t' - t)} +  \mathcal{E}(t') \mathcal{E}^{*}(t) e^{i \omega_{\pm} (t' - t)} \Big )(t'-t) + \mathcal{O}(\phi^2).
\end{align}

\subsection{Effects of the non-linear dispersive shift and Kerr term on the unitary evolution of the qubit-cavity system}
\label{subsec:KerrNonDisperUnitary}

In this section we will show that regardless of the chosen pulse shape (even with numerical tools such as optimal control), to leading order in $\phi_{\pm}$, the changes to the unitary evolution of the qubit-cavity system due to the non-linear dispersive shift and Kerr terms cannot be completely removed for time scales $T \ll \frac{1}{\phi_{\pm}}$. Using \cref{eq:FinalRRbeforExpand,eq:FinalAforReal2,eq:FinalBforReal3}, we can write (for instance choosing the terms affecting the $|0\rangle \langle 0|$
\begin{align}
R_{+}(T)D(A_{+}) = e^{- i \omega_+ T a^{\dagger} a} e^{i \omega_+ T \phi_{+}
  (a^{\dagger} a)^2} e^{I_1 a^{\dagger} - I_1^{*} a - I_2 \phi_{+} (2
  a^{\dagger} a - 1) a^{\dagger} + I_2^{*} \phi_{+} a (2 a^{\dagger} a - 1)},
\end{align}
where we define
\begin{align}
I_1 \equiv  -i \int_{0}^{T} \mathcal{E}(t) e^{i \omega_{+} t}dt,
\end{align}
and
\begin{align}
I_2 \equiv \int_{0}^{T} \mathcal{E}(t)te^{i\omega_{+}t}dt.
\end{align}

Let $A = i \omega_+ T \phi_{+} (a^{\dagger} a)^2$ and $B = I_1 a^{\dagger} -
I_1^{*} a - I_2 \phi_{+} (2 a^{\dagger} a - 1) a^{\dagger} + I_2^{*}
\phi_{+} a (2 a^{\dagger} a - 1)$. Using the Baker-Campbell-Hausdorff lemma and keeping only leading order terms in $\phi_{+}$, we have 
\begin{align}
e^A e^B =  e^{A + B + \frac{1}{2} [A, B] - \frac{1}{12} [B [A, B]] -
  \frac{1}{720} [B, [B, [B, [B, A]]]]}.
\end{align}
Computing the commutators, we find
\begin{align}
e^{A}e^{B} &= e^{\frac{1}{6} i \omega_+ T \phi_{+} \left( - \frac{1}{5} | I_1 |^4 + |
  I_1 |^2 \right)} e^{ \big [ \left( I_2 \phi_{+} + \left( 1 + \frac{1}{2} i \omega_+ T
  \phi_{+} \right) I_1 \right) a^{\dagger} - \left( I_2^{\dagger} \phi_{+} + \left( 1
  - \frac{1}{2} i \omega_+ T \phi_{+} \right) I_1^{\dagger} \right) a +
  \frac{1}{6} i \omega_+ T \phi_{+} (I_1^2 (a^{\dagger})^2 + 4 | I_1 |^2
  a^{\dagger} a + (I^{\dagger})^2 a^2)} \nonumber \\
  & ^{+ i \omega_+ T \phi_{+} (I_1 a^{\dagger} a^{\dagger} a + I_1^{\dagger}
  a^{\dagger} a a) - 2 I_2 \phi_{+} a^{\dagger} a a^{\dagger} + 2 I_2^{\dagger}
  \phi_{+} a a^{\dagger} a + \omega_+ T \phi_{+} (a^{\dagger} a)^2 \big ] }
  \label{eq:LastTermPhi}
\end{align}

Notice that the last term in \cref{eq:LastTermPhi} is proportional to $(a^{\dagger}a)^2$ and independent of the pulse shape (performing the same calculation as above including the $B_+$ term will not change the conclusion). The terms dependent on the pulse shape are expressed as lower powers of $a$ and $a^{\dagger}$. One cannot eliminate all terms proportional to $\phi_{+}$ unless one chooses a time scale comparable to $\frac{1}{\phi_{+}}$. For time scales on the order of $\frac{1}{\phi_{+}}$, higher order terms in $\phi_{+}$ will be relevant and therefore it might possible to choose pulse (with numerical techniques such as optimal control) which can eliminate effects from the non-linear dispersive shift and Kerr terms. However using the parameters of \cref{Tab:ParameterValuesMasterEquation}, $\frac{1}{\phi_{+}}$ is roughly four orders of magnitude longer than the chosen time scale ($T = \frac{\pi}{\chi}$) of our protocol. For such long time scales, effects due to photon loss, damping and dephasing would render the protocol impractical. 

\section{Probability of measuring accepted measurement strings}
\label{APP:ProbabilitySection}

In this section, we derive the analytic expression for output states obtained from phase estimation protocols and their corresponding probabilities. 

\subsection{Phase estimation measurement operator of 1 round} 

From \cref{fig:PhaseEstimationCircuit}, the operator describing the evolution of a single round of phase estimation in terms of the measurement operator set $\Big \{M_{x} = \frac{D(-\sqrt{2\pi}) + (-1)^{x}e^{i\gamma}D(\sqrt{2\pi})}{2} | x = 0,1  \Big \}$ can be written as, 

\begin{align}
    U_{\text{measure}}^\gamma =\Big\{ \ket{0}\bra{0}\otimes M_{0}+\ket{1}\bra{1}\otimes M_{1} \Big\}
\end{align}

After the measurement, the state become,

  \begin{align}
     \ket{\psi_{\text{output}}} &= \frac{U_{\text{measure}}^\gamma\ket{\psi_{\text{input}}}}{\sqrt{\text{Pr}_{x}}}\nonumber\\
     &=\frac{1}{2\sqrt{\text{Pr}_{x}}}\Big( D(-\sqrt{2\pi}) \ket{\psi_{\text{input}}} + (-1)^x e ^{i\gamma}D(\sqrt{2\pi})  \ket{\psi_{\text{input}}}\Big) \nonumber \\
     &=\frac{1}{2\sqrt{\text{Pr}_{x}}}\Big( D(-\sqrt{2\pi}) \ket{\psi_{\text{input}}} + e^{i(\gamma + x \pi)} D(\sqrt{2\pi}) \ket{\psi_{\text{input}}}\Big)
     \end{align}

     where $\text{Pr}_x$ is the probability of measuring $x \in \{0, 1\}$. 

\subsection{Phase estimation measurement operator after $M$ rounds} 
After $M$ arounds of the phase estimation circuit,  the following state will be generated:
\begin{align}
    \ket{\Phi(x[M])} =\frac{1}{2^M\sqrt{N}} \prod_{m=1}^{M}\big[D(-\sqrt{\frac{\pi}{2}})+e^{i(\gamma_m+x_m\pi)}D(\sqrt{\frac{\pi}{2}}) \big]\ket{\text{sq.vac}}
\end{align}
Where $N = \prod_i \text{Pr}_{x_i}$ is the normalization factor. From the product of $M$ sums, if we choose  $j$ of them to be positive shifts ($D(\sqrt{\frac{\pi}{2}})$), $M-j$ to be negative shifts ($D(-\sqrt{\frac{\pi}{2}})$), we can produce a peak at $\sqrt{\pi/2}(j-(M-j))=\sqrt{2\pi}(j -M/2)$ in $q$ space, and the produced phase is  $\prod_{l\in S_j}e^{i(\gamma_l+x_l\pi)}$ ($S_j$ is the set of rounds we choose to be positive shifts). There are $\begin{pmatrix}M\\m\end{pmatrix}$ ways to choose such an $S_j$, by choosing any of such $S_j$ we can produce a peak at $\sqrt{2\pi}(j -M/2)$. Thus, $\ket{\Phi(x[M])}$ can be written as:
\begin{align}
    \ket{\Phi(x[M])}=\frac{1}{2^M \sqrt{N}}\sum_{j=0}^{M}c_j(x[M])D(\sqrt{2\pi}(j-M/2))\ket{\text{sq.vac}} 
    \label{m_round_state}
\end{align}

where $N\approx\sum_{j=0}^{M}|c_j(x[M])|^2$ with $c_j(x[M])=\sum_{\{S_j\}}\prod_{l\in S_j}e^{i(\gamma_l +x_l\pi)}$.

\subsection{Probability} 
Assume we already performed $M-1$ rounds of phase estimation and the measurement results $x_1, ..., x_{M-1}$ and angle parameters we chose (denoted as $\gamma_1,...\gamma_{M-1}$) are known.  Then from \cref{m_round_state}, the input state to the $M$th round is (including the qubit):
\begin{align}
    \ket{\Phi_{input}}_M = \frac{1}{2^{M-1}\sqrt{Pr_{x_1}...Pr_{x_{M-1}}}}\sum_{j=0}^{M-1}c_j(x[M-1])D(\sqrt{2\pi}(j-(M-1)/2))\ket{0}\otimes\ket{\text{sq.vac}}.
\end{align}

The probability of measuring $x_M$ is thus given by
\begin{align}
    Pr_{x_M}&= \bra{ \Phi_{\text{input}}} (U_{\text{measure}}^{\gamma})^{\dagger} U_{\text{measure}}^{\gamma} |  \Phi_{\text{input}} \rangle_{M} \nonumber \\
    &=\Big|\frac{1}{2^{M}\sqrt{Pr_{x_1}...Pr_{x_{M-1}}}}\sum_{j=0}^{M}c_j(x[M])D(\sqrt{2\pi}(j-(M)/2))\ket{\text{sq.vac}}\Big|^2\\ 
\end{align}

Notice that the transfer probability from one peak to another is almost 0 ($<10^{-35}$) assuming an initial squeezing level of 0.2. Therefore, we can make the following approximation
\begin{align}
&\approx \frac{1}{2^{2M}Pr_{x_1}...Pr_{x_{M-1}}} \sum_{j=0}^{M}\Big|c_j(x[M])D(\sqrt{2\pi}(j-(M)/2))\ket{\text{sq.vac}}\Big|^2 \\
    &=\frac{1}{2^{2M}Pr_{x_1}...Pr_{x_{M-1}}}\sum_{j=0}^{M}\Big|c_j(x[M])\Big|^2 \underbrace{\langle{\text{sq.vac}|D^{\dagger}(\sqrt{2\pi}(j-(M)/2))|D(\sqrt{2\pi}(j-(M)/2))|\text{sq.vac}}\rangle}_{1}\\
     &=\frac{1}{2^{2M}Pr_{x_1}...Pr_{x_{M-1}}}\sum_{j=0}^{M}\Big|\sum_{\{S_j\}}\prod_{l\in S_j}e^{i(\gamma_l +x_l\pi)}\Big|^2\\
    &=\frac{1}{2^{2M}Pr_{x_1}...Pr_{x_{M-1}}}\sum_{j=0}^{M}\Big[\Big(\sum_{\{S^m_j\}}\prod_{l\in S^m_j}e^{-i(\gamma_l +x_l\pi)}\Big)\Big(\sum_{\{S^n_j\}}\prod_{k\in S^n_j}e^{i(\gamma_k +x_k\pi)}\Big)\Big]
\end{align}

The above expression is recursive, hence substituting $Pr_{x_{M-1}}$ into it, we find
\begin{align}
    Pr_{x_M}& \approx \frac{\sum_{j=0}^{M}\Big[\Big(\sum_{\{S^m_j\}}\prod_{l\in S^m_j}e^{-i(\gamma_l +x_l\pi)}\Big)\Big(\sum_{\{S^n_j\}}\prod_{k\in S^n_j}e^{i(\gamma_k +x_k\pi)}\Big)\Big]}{4\sum_{j=0}^{M-1}\Big[\Big(\sum_{\{S^m_j\}}\prod_{l\in S^m_j}e^{-i(\gamma_l +x_l\pi)}\Big)\Big(\sum_{\{S^n_j\}}\prod_{k\in S^n_j}e^{i(\gamma_k +x_k\pi)}\Big)\Big]}
\end{align}

\section{Details of numerical simulation}
\label{APP:ProbabilitySection}

\begin{figure}
    \centering
    \includegraphics[height=4.5cm]{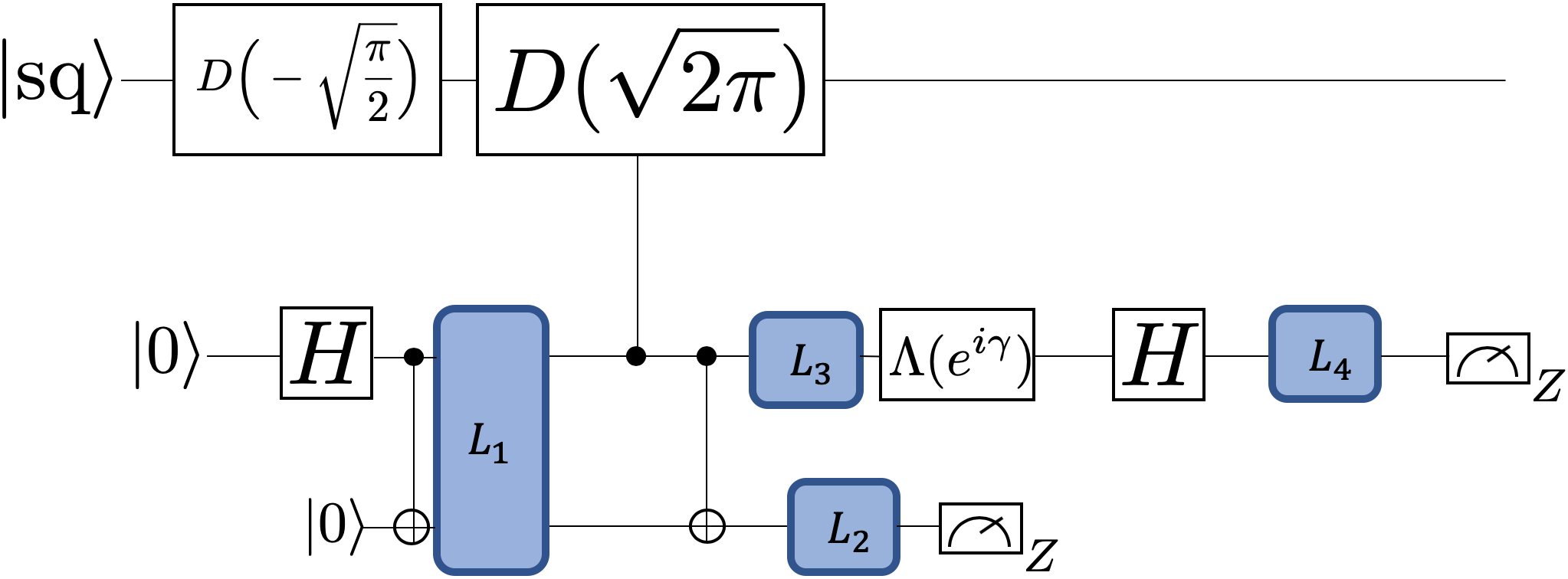}   
    \caption{Illustration of error locations (blue boxes) before and after the controlled displacement gate, where Pauli errors were added based on their corresponding error probability polynomials (computed from the depolarizing channel described in \cref{subsec:MasterEquationAnalysis}). Errors during the controlled-displacement gate were simulated using the master equation described in \cref{subsec:MasterEquationAnalysis}. }
    \label{fig:ErrorLocationsCircuits}
\end{figure}

From the depolarizing noise model described in \cref{subsec:MasterEquationAnalysis}, we computed error probabilities for all Pauli errors arising from a single fault at the locations indicated by the blue boxes in \cref{fig:ErrorLocationsCircuits}. For instance, the probability of an $I \otimes Z$ error at location $L_1$ is given by
\begin{align}
P_{IZ} = \frac{p}{15}\Big (  P1_{X}P1_{Z} + P1_{X}(1-P1_{Z}) + (1-P1_{X})(1-P1_{Z}) \Big) + (1-p)(1-P1_{X})P1_{Z},
\end{align}
where
\begin{align}
P1_{X} = \frac{p}{15},
\end{align}
during the first round, and 
\begin{align}
P1_{X} =\frac{2p}{3},
\end{align}
in later rounds. 
Similarly,
\begin{align}
P1_{Z} = \frac{p}{15}(1-\frac{p}{10}) + \frac{p^2}{450} + (1-\frac{p}{15})\frac{p}{15},
\end{align}
for the first round and 
\begin{align}
P1_{Z} = \frac{2p}{3}(1-\frac{p}{10}) + \frac{2p}{3}\frac{p}{30} + (1-\frac{2p}{3})\frac{p}{15}.
\end{align}

With a Pauli error added at either locations $L_1$, $L_2$, $L_3$ or $L_4$, we update the state of the qubit before performing the master equation simulation described in  \cref{subsec:MasterEquationAnalysis}. With the analytic error probability for the Pauli error, in addition to the evolution of the state during the master equation, we can compute the total probability of obtaining the output state. As mentioned in \cref{subsec:MasterEquationAnalysis}, such an analysis ignored second order Pauli error events (before and after the controlled displacement gate).

\end{document}